\documentclass[useAMS,usenatbib,onecolumn]{mn2e}
\usepackage{times}
\usepackage{graphicx}
\usepackage{natbib}
\usepackage{bm}
\usepackage{epstopdf}
\usepackage{color}

\def\asca{{\sl ASCA }}

\def\xmm{{\sl XMM-Newton }}
\def\chandra{{\sl Chandra }}
\def\fuse{{\sl FUSE }}

\def\fek{Fe~{\sc k}}
\def\fexxv{Fe~{\sc xxv}}

\def\fexxiii{Fe~{\sc xxiii}}
\def\fexxii{Fe~{\sc xxii}}
\def\fexxi{Fe~{\sc xxi}}
\def\fexx{Fe~{\sc xx}}
\def\fexix{Fe~{\sc xix}}
\def\fexviii{Fe~{\sc xviii}}
\def\fexvii{Fe~{\sc xvii}}
\def\fex{Fe~{\sc x}}

\def\fexxib{Fe~{\sc xxi-xxiii}}
\def\feib{Fe~{\sc i-xii}}

\def\nvi{N~{\sc vi}}
\def\nvii{N~{\sc vii}}

\def\cv{C~{\sc v}}
\def\cvi{C~{\sc vi}}

\def\neviii{Ne~{\sc viii}}
\def\neix{Ne~{\sc ix}}
\def\nex{Ne~{\sc x}}

\def\oviii{O~{\sc viii}}
\def\ovii{O~{\sc vii}}
\def\ovi{O~{\sc vi}}
\def\oi{O~{\sc i}}

\def\mgxi{Mg~{\sc xi}}
\def\mgxii{Mg~{\sc xii}}
\def\sivii{Si~{\sc vii}}
\def\siviii{Si~{\sc viii}}
\def\siix{Si~{\sc ix}}
\def\six{Si~{\sc x}}
\def\sixi{Si~{\sc xi}}
\def\sixiii{Si~{\sc xiii}}
\def\sixiv{Si~{\sc xiv}}

\def\sxv{S~{\sc xv}}

\title[Ionized Absorbing Gas in Mrk 290]{\chandra and \xmm view of the Warm Absorbing Gas in Mrk 290}
\author[Zhang et al.]{ Shuinai Zhang$^{1}$\thanks{E-mail:
snzhang@nju.edu.cn}, L. Ji$^2$, H. L. Marshall$^2$, A. L. Longinotti$^2$, D. Evans$^2$, Q. S. Gu$^{1,3}$\\
$^{1}$Department of Astronomy, Nanjing University, Nanjing 210093, China\\
$^{2}$MIT Kavli Institute, Cambridge, MA 02139, U.S.A.\\
$^3$Key Laboratory of Modern Astronomy and Astrophysics (Nanjing University), Ministry of Education, China}

\begin{document}

\date{Accepted ??; Received ??; in original form ??}

\pagerange{\pageref{firstpage}--\pageref{lastpage}} \pubyear{2010}

\maketitle

\label{firstpage}

\begin{abstract}
We present a detailed analysis of the \chandra High Energy Transmission Grating Spectrometer (HETGS) 
and \xmm high resolution spectra of the bright Seyfert 1 galaxy, Mrk 290.  

The \chandra HETGS spectra reveal complex absorption features that can be best described by a combination of three ionized absorbers.
The outflow velocities of these warm absorbers are about $450\,\rm{km\,s^{-1}}$, consistent with the three absorption components found in a
previous far UV study.
The ionizing continuum of Mrk 290 fluctuated by a factor of 1.4 during \chandra observations
on a time scale of 17 days. 
Using the response in opacity of the three absorbers to this fluctuation, 
we put a lower limit on the distance from the ionizing source of 0.9 pc for the medium ionized absorber and an upper limit on distance of 2.5 pc for the lowest ionized absorber.
The three ionization components lie on the stable branch of the thermal equilibrium curve, indicating roughly the same gas pressure.
Therefore the thermal wind from the torus is most likely the origin of warm absorbing gas in Mrk 290.

During the \xmm observation, the ionizing luminosity was 50\% lower compared to the one in the \chandra observation.
The RGS spectrum is well fitted by a two phase warm absorber, with several additional absorption 
lines attributed to a Galactic high velocity cloud, Complex C.
Neither the ionization parameter $\xi$ nor the column density $\rm{N_H}$ of the two absorbing components varied significantly, compared to the results from \chandra observations. 
The outflow velocities of both components were $1260\,\rm{km\,s^{-1}}$. We suggest that an entirely new warm absorber from the torus passed through our line of sight.

Assuming the torus wind model, the estimated mass outflow rate is $\sim1M_{\odot}$ per year, while the nuclear accretion rate is $\sim0.5M_{\odot}$ per year.
The \ovii~and \neix~forbidden lines are the most prominent soft X-ray emission lines, 
with a mean redshift of $700\,\mathrm{km\,s^{-1}}$ relative to the systematic velocity.
There seems to be no relation between emission lines and warm absorbers.

\end{abstract}

\clearpage

%%%%%%%%%%%%%%%%%%%%%%%%%%%%%%%%%%%%%%%%%%%%%%%%%%%%%%%%%%%% 
\section{Introduction}
Ionized outflowing gas (or `warm absorber', WA) is believed to be common in the inner parts of Active Galactic Nuclei (AGN) based on X-ray spectroscopy. 
About 50\% of  Seyfert 1s \citep{reynolds97, george98} and quasars \citep{piconcelli05, misawa07, ganguly08} are characterized by blue-shifted X-ray absorption lines, the velocities of which are usually several hundred $\rm{km\,s^{-1}}$. 
UV narrow absorption lines from outflowing gas were also detected at a similar fraction \citep{crenshaw99}, and showed velocities and ionization states as similar to WAs in some cases \citep{gabel05a, gabel05b, costantini07}. 
However, in other cases there is no clear physical correspondence between them \citep{kriss04}.

Several studies have shown that WAs can be well modeled by including multiple absorbing components, which appear to be in pressure equilibrium \citep[e.g.][]{netzer03, mckernan07, krongold09}.
Thus, the structure of WAs can be interpreted as a multi-phase medium, and most likely occupies the same spatial region \citep{krolik01}.
Another suggestion that WAs occur in continuous density-stratified streamers \citep{behar03, steenbrugge05} seems to be ruled out \citep{krongold05, krongold07}.

Unfortunately, very little is still known about the physical states of WAs, or their dynamics and geometry. The most fundamental question is where WAs originate.
Proposed locations span a wide range in radial distance from the central ionizing source: the accretion disk \citep{elvis00, krongold07}, the broad line region (BLR) \citep{kraemer05}, the putative obscuring torus \citep{krolik01,blustin05} and the narrow line region \citep{behar03, crenshaw09}.
Simulations have offered similar possibilities, such as the accretion disk wind \citep{proga04, risaliti09}, the wind from torus \citep{dorodnitsyn08}, the large-scale outflows \citep{kurosawa09}, but ruled out few options.

In principle, these outflows could potentially provide a common form of AGN feedback.
But whether the outflow winds could extend into the interstellar medium and intergalactic medium \citep{nicastro08} or just fall back into the accretion disk \citep{stoll09},
depends critically on the exact location of WAs when estimating the mass outflow rate and the total outflow kinetic energy. 
One key diagnostic for the location of WAs is based on applying non-equilibrium models on time variable AGNs \citep{nicastro99}.
Such strong variability needed to apply to this method is not that common.

Mrk 290, a relatively nearby Seyfert 1 galaxy ($z=0.0304\pm0.0014$ from the Sloan Digital Sky Survey), has been extensively  studied by low resolution instruments \citep[e.g.][]{wood84, kruper90}. The WA was found in the \asca observation \citep{turner96}. The source flux of 2-10 keV band was $\rm{8.4\times10^{-11}\,erg\,s^{-1}\,cm^{-2}}$ in the \asca observation.
In 2003, the first high resolution observation was carried out using \chandra HETGS, simultaneously with Far Ultraviolet Spectroscopic Explorer (\fuse) to determine the ionization state and kinematics of the WA, and assess association with the UV-absorbing gas. Mrk 290 was observed by  \xmm in 2006.

We present the analysis of the ionized absorbers in Mrk 290, using the high resolution X-ray spectra.
This paper is organized as follows. In section 2, we describe the data reduction. Section 3 is devoted to the analysis of the data, and section 4 to the detailed model of the ionized absorbers of Mrk 290. In section 5 we discuss our results and present our conclusions in section 6. 
The cosmological parameters used are: $\mathrm{H_0=73\,km}$ $\mathrm{s^{-1}\,Mpc^{-1}}$,
$\Omega_\Lambda=0.72$ and $\Omega_m=0.24$ \citep{spergel07}. The luminosity distance is 123.7 Mpc in this cosmological model.
The quoted errors refer to 90\% confidence for one interesting parameter.

%%%%%%%%%%%%%%%%%%%%%%%%%%%%%%%%%%%%%%%%%%%%%%%%%%%%%%%%%%%% 
\section{Observation and Data Reduction} 

%******************************************************************
\subsection{\chandra Data: 2 Flux States}

\chandra performed four observations of Mrk 290 for a total duration of 251 ks, using the High Energy Transmission Grating Spectrometer \citep[HETGS,][]{canizares05}. Table~\ref{tab:id} lists the observation log. 

The HETGS is composing of two grating sets, the High Energy Gratings (HEG) and the Medium Energy Gratings (MEG), the resolutions of which are 0.012 \AA ~and 0.023 \AA, respectively. 
All observations were reduced uniformly and in a standard way using the \chandra Interactive Analysis of Observations (CIAO) software (Version 4.1.1). Flux calibration of each observation was carried out using the \chandra Calibration Database (Version 4.1.2).  The +1st and -1st orders of each HEG and MEG spectra were combined; higher orders were ignored.

\begin{figure*} %  figure placement: here, top, bottom, or page
   \centering
   \includegraphics[angle=-90,width=2.5in]{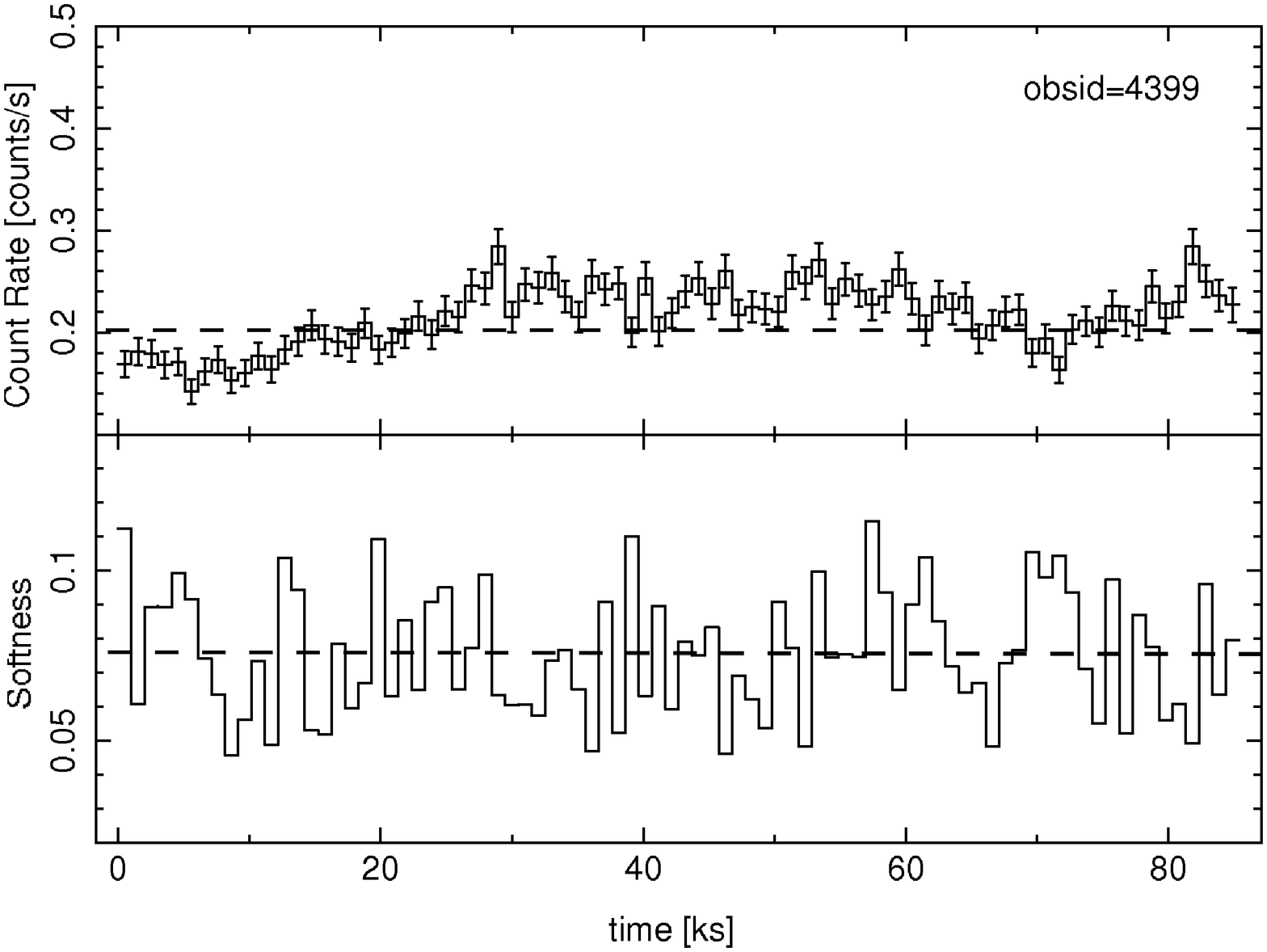}  
      \includegraphics[angle=-90,width=2.5in]{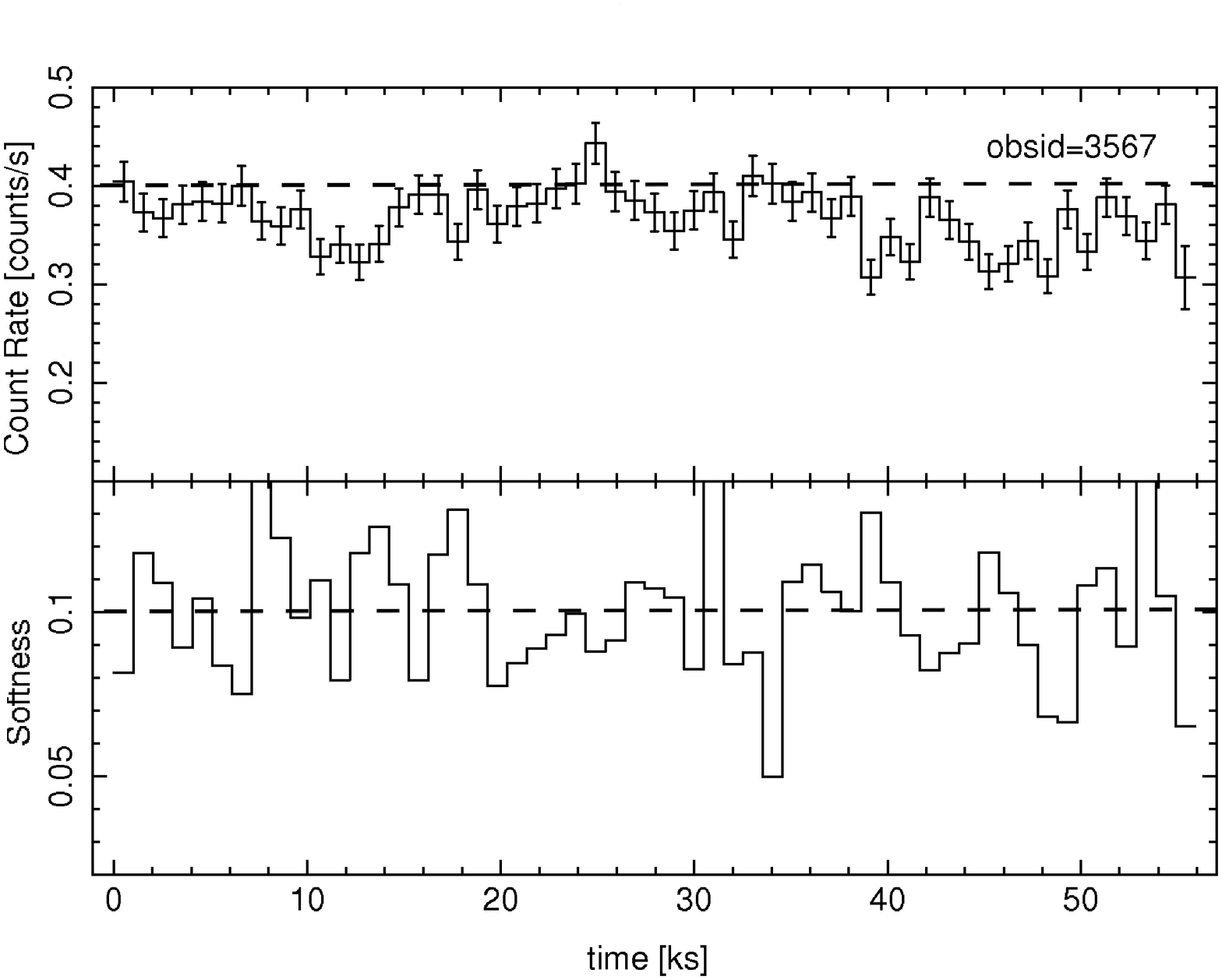} 
         \includegraphics[angle=-90,width=2.53in]{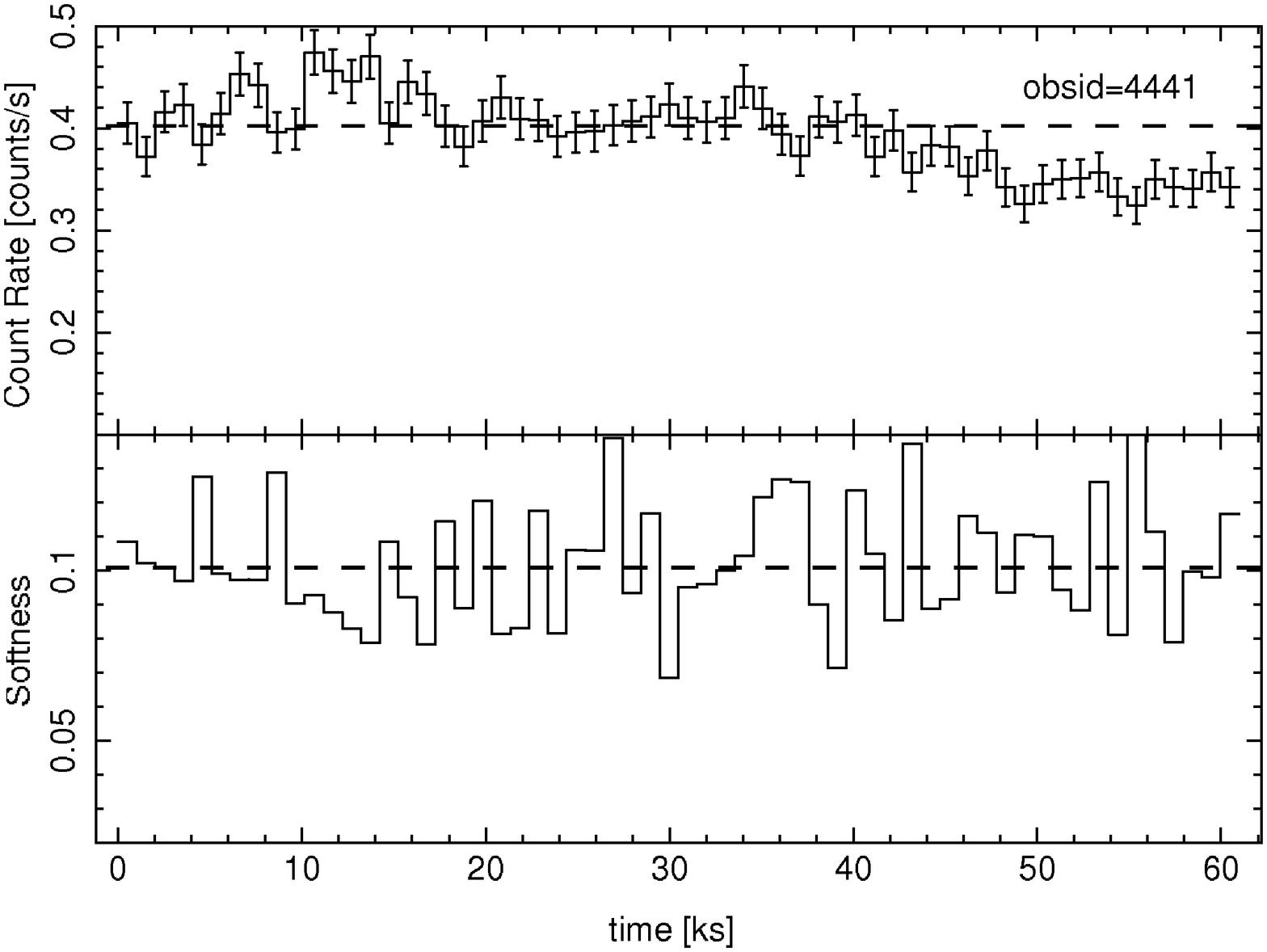} 
            \includegraphics[angle=-90,width=2.53in]{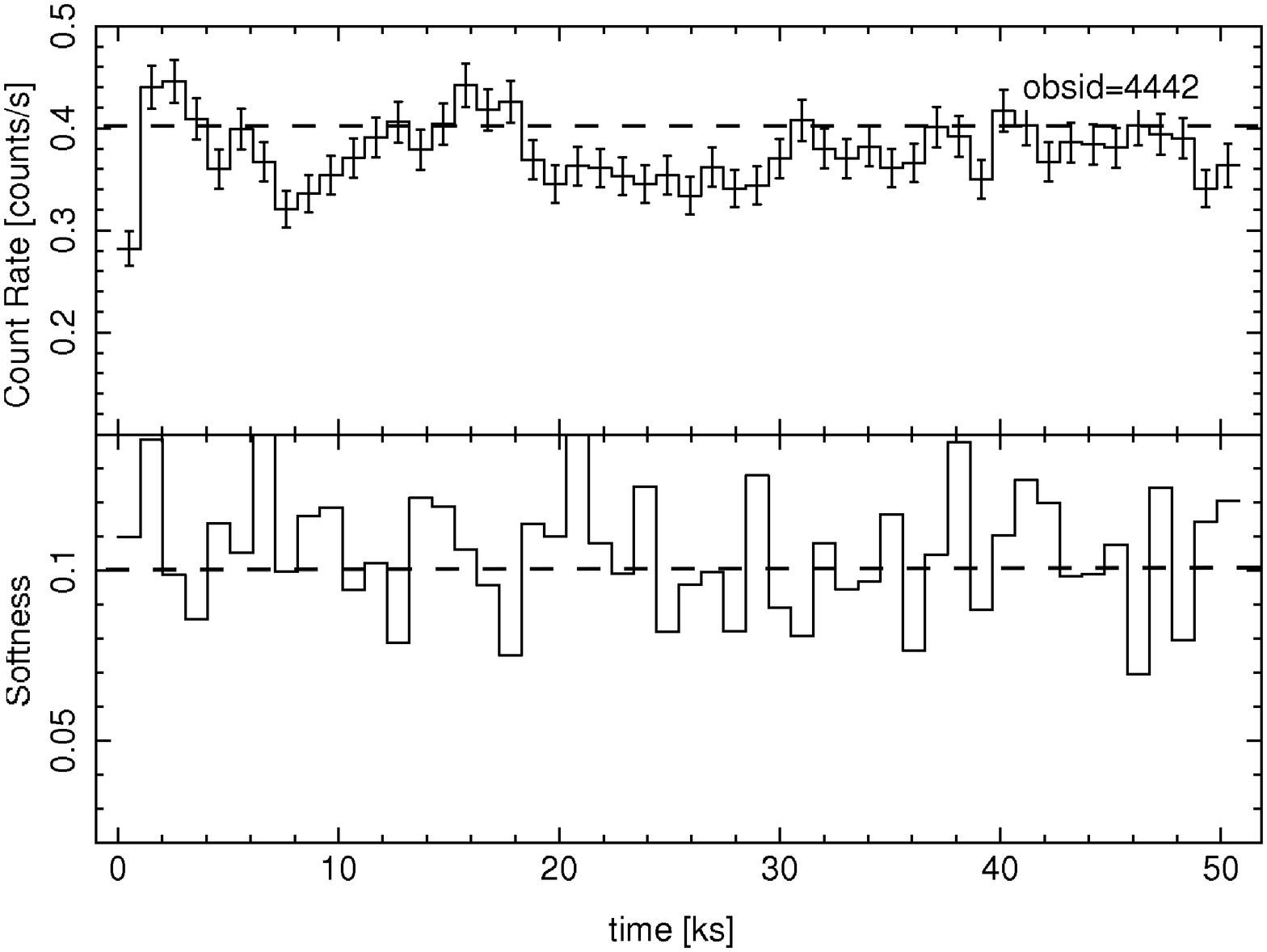} 
   \caption{Light curves from \chandra observations of Mrk 290. Each panel corresponds to an observation in Table 1. The upper half of each panel shows the hard X-ray (2-10 \AA) count rate and the lower half shows the softness ratio, defined as the count rate in the 15-25 \AA~band divided by that in the 2-10 \AA. The dashed lines serve as variability references. The count rate and softness ratio are lowest in observation 4399, while the other three observations are consistent with each other.}
   \label{fig:lc}
\end{figure*}

We generated the light curves in 1000 second bins for the short wavelength band (2-10 \AA), as shown in Fig.~\ref{fig:lc}. The comparison between them indicates that the time-average count rate increased by a factor of $\sim$ 2, between observations 4399 and 3567, only half a month apart. 
Smaller variations are observed on time scales of $10^4$ s within each observation, which were also found by \citet{turner96}.
Figure~\ref{fig:lc} shows the softness ratio, defined as the count rate in the 15-25 \AA ~band divided by that in the 2-10 \AA~band, to illustrate the different variability characteristics below and above 1 keV \citep{netzer03}. The softness ratio increases from 0.075 to 0.100 along with a higher count rate, between observations 4399 and 3567.
In light of these results, we have divided the entire dataset into two groups with higher (ID 3567, 4441 \& 4442) and lower (ID 4399) count rates.

%******************************************************************
\subsection{\xmm Data Reduction}

Mrk 290 was observed four times by  \xmm \citep{jansen01} during the period April 30 to May 6, 2006.
The European Photon Imaging Camera (EPIC) was operated in small window mode. Both RGS (Reflection Grating Spectrometer) cameras were in spectroscopic mode. 
Raw data from all observations were processed with the standard Science Analysis System (SAS) using version 8.0 and the most updated calibration files.

High `flaring' background periods were all filtered to maximize the signal-to-noise ratio (S/N). 
While the total exposure time is 100 ks, after this screening process, the net exposure time was 65.5 ks, as shown in Table~\ref{tab:idxmm}. 
The four \xmm data were merged, since no significant flux variability was found.

The {\it pn} spectra were extracted using pattern 0-4. 
The 0.5-10 keV count rate, 2.5 counts s$^{-1}$, was under the pile-up threshold, verified with the {\footnotesize EPATPLOT}.
Each source spectrum was extracted from a circular region of 40$''$ of radius centered on the peak of the X-ray emission.
The background spectrum was taken from a vicinity source-free circular region of radius of 55$''$.
RGS data were processed with the task {\footnotesize RGSPROC} according to the standard method proposed in the SAS thread.

%%%%%%%%%%%%%%%%%%%%%%%%%%%%%%%%%%%%%%%%%%%%%%%%%%%%%%%%%%%% 
\section{Spectral Analysis}

We used the Interactive Spectral Interpretation System (ISIS version 1.5.0, \citep{houck02}) to fit the spectra.
When fitting the data, we used the $C$ statistic \citep{cash76} to find the best-fitting model parameters in the high-resolution spectra.

%******************************************************************
\subsection{The Broadband Continuum Modeling}

The Mrk 290 spectrum is affected by a relatively low Galactic absorption: $N_{H}=1.76\times10^{20}\,\mathrm{cm}^{-2}$ \citep{kalberla05}, which is included in all  models.

The \chandra spectra were binned to 0.01 \AA, and fitted in the wavelength range 1.5-25 \AA, as shown in Fig.~\ref{fig:none1}.
In contrast to previous results, where Mrk 290 was described by an absorbed power law \citep{turner96, shinozaki06, jiang06},
we found that  a black-body component was needed in addition to the absorbed power law in order to fit the continuum.
The black-body component was included to account for the soft excess, that could come from a standard optically-thick accretion disk.
The intrinsic cold absorption was negligible compared to the Galactic column density.
In fact, half of AGN \citep{mckernan07} and quasars \citep{piconcelli05} do not appear to have significant intrinsic cold gas absorption.

The {\it pn} spectrum was grouped to give a minimum of 50 counts  per bin in order to apply $\chi^2$ statistics. 
The same model as fitted to HETGS spectra but with free parameters provided a good fit to the {\it pn} data, except for two prominent residuals around 0.8 and 6.3 keV.
These two features are consistent with typical absorbing features of WA and \fek$\alpha$ emission line, respectively.
Again, no cold absorption component above Galactic was needed.
Fig.~\ref{fig:pn} shows the spectrum in the 0.5-10 keV band.

The best fit parameters of all spectra are given in Table~\ref{tab:conti}, including the fluxes and luminosities in the 2-10 keV band.
The hard X-ray luminosity increased by a factor of 1.4 in 17 days between \chandra observations. 
When \xmm re-observed the source three years later, the luminosity showed a decrease of 50\% compared to the highest value measured in the \chandra spectra.
So hereafter, we refer to these spectra as `high-', `mid-' and `low-state' spectra as labeled in Table~\ref{tab:conti}.

 %******************************************************************
\subsection{Intrinsic Absorption Line Properties}
The HETGS spectra, which are binned to 0.01 \AA, reveal many absorption line features, some of which are blends of several lines.
H-like and He-like lines of O, Ne, Mg, Si, S, and L-shell lines of Si, Fe are marked in Fig.~\ref{fig:none1}.
The M-shell complex of iron, called the unresolved transition array (UTA) is observed as a depression at 16-17.5 \AA, although it is very shallow in the present data.
We identified one prominent narrow absorption line of \fex~in the UTA (see Table~\ref{Tab:habs}). 
All statistically significant lines with Poisson probability greater than 0.995 were measured by fitting a negative Gaussian to the continua within $\pm0.5\,$\AA~band, as shown in Table~\ref{Tab:habs}. 
The widths of the Gaussian lines were all fixed to 1 m\AA.
The outflow velocities about $450\,\rm{km\,s^{-1}}$ were derived with respect to the systemic redshift 0.0304, except those lines labeled as Galactic absorption.There are fewer significant lines in the mid-state spectrum than in the high-state spectrum.

The RGS spectra are binned to 0.1 \AA, in which the iron UTA shows clearly.
We fitted RGS 1 and 2 simultaneously in the wavelength band 7-35 \AA, but for clarity added them in Fig.~\ref{fig:rgsc}.
The continuum model was fixed to that obtained from {\it pn} spectrum, except for allowing the black body component to vary.
We found several highly ionized features: \neix ~combined with \fexix, \nex ~Ly$\alpha$, \cvi, \nvii, \fexviii ~and \fexx~complex.
A Gaussian was fitted to each feature and the results were listed in Table~\ref{Tab:habs}.
The outflow velocities of these lines are more than 1000 $\rm{km\,s^{-1}}$, differing by the several hundred $\rm{km\,s^{-1}}$ from those derived from the HETGS spectra.

%**************************************************************************
\subsection{Galactic Absorption}
The spectra of Mrk 290 show narrow absorption lines at zero redshift in Table~\ref{Tab:habs}.
\oi~absorption at 23.5 \AA~in Fig.~\ref{fig:rgsc} is attributed to the ISM of our Galaxy \citep{wilms00}.
We also detected ionized material traced by \oviii~in the HETGS spectra, and \ovii, \nvii, \nvi, and possibly \cv~in the RGS spectrum.
This ionized absorber may be associated with the high-velocity cloud complex C, which is a low-metallicity gas cloud plunging toward the disk and beginning to interact with the ambient gas that surrounds the Milky Way \citep{collins07}. In the UV band, \citet{penton00} detected absorption lines from complex C using HST observations.
Since this ionized gas does not affect the continuum shape \citep{costantini07}, we fitted these lines using Gaussians. 
The upper limits of the infall velocity with respect to our Milky Way were measured from these lines less than $\rm{1000\;km\,s^{-1}}$, 
consistent with the values from the UV lines: $\rm{200\;km\,s^{-1}}$ and $\rm{120\;km\,s^{-1}}$ \citep{collins07}.

%******************************************************************
\subsection{Emission Lines}
In the soft X-ray band, Mrk 290 exhibits a prominent \neix~forbidden line in the high-state spectrum and a \ovii~forbidden line in the high- and low-state spectra. 
They are red-shifted at offset velocities of $700\,\rm{km\,s^{-1}}$ with respect to the systemic redshift, which is definitely different from the absorption lines.
The velocities do not change over three years suggests that these emission lines may be emitted in the same gas.
The fit results for the three lines are shown in Table~\ref{Tab:emit}.
%The \ovii~forbidden line is also detected in the \chandra spectra at low significance (Fig.~\ref{fig:none1}), at a similar redshift. 

The iron emission line at 6.4 keV (or 1.9378 \AA) is present in both the high-state HETGS spectrum and the low-state {\it pn} spectrum. 
This fluorescent line is K$\alpha$ transitions from neutral iron, consistent with \feib. 
A Gaussian line, the width of which was fixed at 1 m\AA, had been included to fit the \fek$\alpha$ line in the 5-8 keV band.
Table~\ref{Tab:emit} lists the measurements.
The equivalent widths (EWs) of $48\pm17$ eV for the high-state spectrum and of $23\pm17$ eV for the low-state spectrum are consistent with the result of \citet{shu10}. 
However they are much smaller than the EWs from \asca data: 273 eV \citep{shinozaki06} or 500 eV \citep{turner96}.

We allowed the line width to float and fitted the iron K$\alpha$ again.
The Cash statistics changed by 2 and 8 when reducing one $d.o.f.$ for the high- and low-state spectra, respectively, and the equivalent widths (EWs) increased to $71\pm35$ eV and $81\pm29$ eV.
The FWHM of \fek$\alpha$ in the high-state spectrum is $\rm{5700^{+9100}_{-4600}\,km\,s^{-1}}$, and in the low-state {\it pn} spectrum is $\rm{14000^{+10000}_{-6000}\,km\,s^{-1}}$, which agrees with \asca result \citep{turner96}. 

In the high-state spectrum, a line at 1.897 \AA~seems consistent with the n=2-1 resonance transition of \fexxv~(at a rest energy of 6.696 keV). 
This line could be produced in Compton-thin, low velocity photoionized material \citep{bianchi05}. 

There are unidentified emission features with Poisson probabilities less than 0.001 in the high-state spectrum at 2.666 \AA~and 8.725 \AA, of which the fluxs are about $few\times10^6\,\rm{ph\,cm^{-2}\,s^{-1}}$.
However, this two features are probably spurious.
%These  may be from the ion of Ca and some highly ionized Fe. 

%%%%%%%%%%%%%%%%%%%%%%%%%%%%%%%%%%%%%%%%%%%%%%%%%%%%%%%%%%%% 
\section{Photoionization modeling}
The publicly available photoionization code XSTAR\footnote{http://heasarc.gsfc.nasa.gov/docs/software/xstar/xstar.html} is the tool we used to model the physical conditions of the absorbing gas. It includes H- and He-like transitions of most ions, L-shell transitions of Fe, and important M-shell transitions of Fe that contribute to the UTA feature.  
We used PVM\_XSTAR \citep{noble09} to generate grids of models, which provides parallel execution of XSTAR 2.1ln11.

We adopted the default solar abundances \citep{grevesse96} in XSTAR, and used a velocity turbulence ({\it b}-value) of $\rm{170\,km\,s^{-1}}$, which corresponds approximately to the limiting MEG spectral resolution ($\sim300\,\rm{km\,s^{-1}}$ FWHM at 0.5 keV) \citep{mckernan07}. 
This choice is justified by the absence of detectable line broadening.

The photoionization model used here contains three parameters: red-shift {\it z}; total neutral hydrogen column density $N_H$; and the ionization parameter
$\xi=L_{ion}/(n_eR^2)$, where $L_{ion}$ is the ionizing luminosity in the range 1-1000 Ryd, $n_e$  is the electron density, and $R$ is the distance of the ionized gas from the central ionizing source.

%******************************************************************
\subsection{The Spectral Energy Distribution}
First of all, we constructed the spectral energy distributions (SEDs) of Mrk 290 to generate the photoionization models.
However, the intrinsic X-ray spectrum, an important part of the SED, is obtained by fitting the X-ray data with photoionization models.
Thus a two-step iterative process is employed to set the intrinsic spectrum, as described in \citet{yaqoob03}.

In the first step, the ionizing X-ray spectrum is obtained by fitting the continuum without a WA to construct a SED.
The SED is then used to generate a grid of warm absorption models that are fit to the spectra.
The second step is similar to the first step: using the new intrinsic X-ray spectrum for a new SED, which is in turn used to generate new set of photoionization models.

The \chandra mid-state data (ID 4399) was observed simultaneously in the UV band by \fuse.
We adopted this UV data point \citep{dunn08}, while other data were obtained from NASA/IPAC Extragalactic Database (NED): radio data from VLA, 25 $\mu$ data from IRAS, and r band data from SDSS.

For each state, the flux point at 0.5 keV band on the intrinsic X-ray continuum was simply linked to the UV point and to other data in the SED by straight lines in log-log space, as shown in Fig~\ref{fig:sed}.
We derived luminosities of 0.5-13.6 keV band from the intrinsic X-ray spectral models, measured luminosities of 13.6 eV - 0.5 keV band from the two point lines in SEDs, and added them to determine $L_{ion}$.
Detailed intrinsic continuum parameters and luminosities are listed in Table~\ref{Tab:warmabs}.

%******************************************************************
\subsection{Warm Absorbers from Model Results}
\subsubsection{One WA Component}
\citet{turner96} modeled the \asca spectrum of Mrk 290 with one WA component generated using XSTAR, obtaining an ionization parameter $log(\xi)=1.38$ and a column density $N_{H}$ of $8\times10^{21}\,\rm{cm}^{-2}$.
Similarly, we began with one WA in fitting \chandra HETGS spectra. 
The \chandra spectra were binned to 0.01 \AA, and fitted with one ionization component.
The model represents most absorption lines well, and the best fit results are shown in Table~\ref{Tab:onewa}.
The absorber is blue-shifted by $-450\pm30\,\rm{km\,s^{-1}}$ relative to the systemic redshift, and neither its column density nor ionization parameter changed apparently from mid-state to high-state during half a month.
Since the UTA feature is quite clear in the RGS spectrum, one WA component fitting is not suitable for the low-state data.

\subsubsection{Two Phase Warm Absorber}
Coarsely binned high- and mid-state spectra are shown in Fig.~\ref{fig:twohigh}, in which the green lines represent the best fitting model from Table~\ref{Tab:onewa}.
The visual inspection of the spectra reveals the unfitted UTA feature clearly in the 16-17.5 \AA~band, suggesting that the single WA component is not sufficient to account for all absorbing features.

The UTA is a broad feature, and its shape is sensitive to the charge states of Fe producing it, so we rebinned the spectra heavily.
Since the UTA is a key diagnostic of material in a low-ionization state, a second WA with a lower ionization level is required to be included in the model.
We modeled the high-state spectrum using a two phase WA with parameters set free, except the redshift of the higher ionization component was fixed to $z=0.0289$ from Table~\ref{Tab:onewa}.
The two phase warm absorber provides a good fit to the high-state spectrum as shown in Fig.~\ref{fig:twohigh} by the red line model.
The Fe UTA is well fitted by a lower ionized component with $log(\xi)=1.62\pm0.15$, $N_{H}=5.4(\pm1.8)\times10^{20}\,\rm{cm}^{-2}$.
The outflow velocity is similar to that of the high ionization component, and the other best fit parameters are reported in Table~\ref{Tab:warmabs}.

Providing that no properties in Table~\ref{Tab:warmabs} varied in 17 days and the dynamical timescale is much longer than the radiative recombination timescale,
when fitting the mid-state spectrum, we fixed the redshift and column density of each absorbing component to that of high-state, but allowed the ionization parameter to vary. 
Unfortunately, the large uncertainties in $\xi$ of the low ionization component cannot constrain the possible response of opacity to the continuum variation.

The two phase WA model was also used to fit the RGS spectrum, adding several gaussian lines to account for the local absorption lines and the \ovii~forbidden emission line. 
Fig.~\ref{fig:rgsflux} shows the best model for RGS spectrum, and fitting results are listed in Table~\ref{Tab:warmabs}.
As expected, the deep UTA feature comes from the lower ionized material, while other features identified as transitions of \neix, \nex, \fexviii, \fexx~and \nvii, come from the higher ionization component. The \cvi~absorption lines are believed to be associated with the low ionization component \citep{sako01}, though the observed lines are much deeper than what the photoionization model predicts here.
However, it could also be the result from carbon-rich gas. 
The \fexviii~complex is not well fitted here, as in HETGS spectra (as shown in Fig.~\ref{fig:twohigh}), which is partly due to the uncertainties of di-electronic recombination rates for iron L-shell ions \citep{gu03}, and partly due to the uncertainty of the iron abundance (solar abundance is used in the photoionization models).
Neither the ionization parameter nor the column density of the two WA components varied significantly compared to that found in the HETGS spectra, but the outflow velocity increased by a factor of about 2.5.
%However, if we consider the results meticulously, we found the column density $N_H$ of the lower ionized WA increased a little bit, while that of the higher ionized WA decreased slightly.

\subsubsection{The Third, Hottest Ionization Component}
We have applied the two phase warm absorber model back to the 0.01 \AA~binned high-state spectrum again and checked residuals carefully.
Three important Fe L-shell absorption lines were left unfitted, \fexxi~(12.284 \AA), \fexxii~(11.770 \AA), \fexxiii~(10.981 \AA).
A hotter high ionization component is required to explain these absorption lines, as labeled in Fig.~\ref{fig:model}. 
Such a very high ionization absorbing component has been reported in a few sources, e.g. NGC 3783 \citep{netzer03}, NGC 5548 \citep{steenbrugge05} and NGC 985 \citep{krongold09}.

We added the third ionized absorber in the high resolution high-state spectrum to test for the possible presence of this component.
The lowest ionization component was fixed to the values of the two phase WA model fit, where they were best determined.
The fit parameters are given in Table~\ref{Tab:triwa}.
The Cash statistic gives a difference of 45 for adding 3 d.o.f (2590/2341) compared to the model with one WA.   
The model of three ionization components are shown in Fig.~\ref{fig:model}.
The comparison of the best-fitting model to the high-state spectrum can be seen in Fig.~\ref{fig:spec1}, in which the flux spectrum is binned to 0.025 \AA~for clarify.

%******************************************************************
\subsection{The L-shell silicon lines}
In \chandra HETGS spectra, Si L-shell absorption lines do not exist in the fitted model, because so far they have mostly been omitted from the XSTAR database \citep{mckernan07}.
Fig.~\ref{fig:silicon} shows these unfitted lines in the mid- and high-state spectra.
The mid-state spectrum shows \sixi, \siix, and one significant line (at 7.23 \AA) that could be \sivii~with an outflow velocity $2000\,\rm{km\,s^{-1}}$, \siviii~with an inflow velocity $700\,\rm{km\,s^{-1}}$, or spurious. 
In the high-state spectrum \sixi, \six, and \siix~are evident.

The 7.23 \AA~line disappeared totally while \siix~got stronger from the mid-state to the high-state. 
Furthermore the ratio of \siix~to \sixi~changed from $0.6\pm0.5$ to $1.6\pm0.4$ over the two states.
\citet{netzer04} suggested that the Si L-shell absorption lines should be generated from the same ionization component producing the Fe UTA. 
\citet{gu06} calculated the new Fe M-shell data and confirmed this suggestion, especially for the NGC 3783.
The variation of Si L-shell lines in Mrk 290 demonstrated the change of the ionization parameter in the lowest ionization component.

%%%%%%%%%%%%%%%%%%%%%%%%%%%%%%%%%%%%%%%%%%%%%%%%%%%%%%%%%%%% 
\section{Discussion}

%******************************************************************
\subsection{The Location of The WAs}
When the ionizing luminosity $L_{ion}$ and the ionization parameter $\xi$ are known, one would expect to constrain the distance $R$ by estimating the electron density $n_{e}$, since $\xi=L_{ion}/(n_{e}R^{2})$ . 
Studying the variation based on non-equilibrium models is one method to constrain $n_e$ \citep{nicastro99}.
An excellent approximation for the time to reach a new photoionization equilibrium is

\begin{equation}
t_{eq}^{x{^i},x^{i+1}} \sim
\left[ \frac{1}{\alpha_{rec}(x^i, T_{e})_{eq}~n_e} \right] \times\left[ \frac{1}{[\alpha_{rec}(x^{i-1}, T_e)/\alpha_{rec}(x^i, T_{e})]_{eq}
    + [n_{x^{i+1}}/n_{x^{i}}]}\right],
\end{equation}
where $x_i$ is the relative density of ion $i$ of the element $X$, and the recombination time $\alpha_{rec}(x^i, T_{e})$ can be obtained from Shull \& van Steenberg (1982) data.

During \chandra observations, the luminosity increased by a factor of 1.4, thus the log($\xi$) was expected to increase 0.15 dex.
However, the ionization parameter of the medium ionization component, which is best constrained in the \chandra spectra, did not change (Table~\ref{Tab:onewa}). 
Providing the flux increase happened between June 29 and July 15 in 2003 and the high-state observation lasted for 2.5 days, the timescale for the gas to reach the new photoionization equilibrium must be longer than 2.5 days.
Considering the ions \ovi~- \oviii~and \neviii~- \nex, we find the electron density of the medium ionization component $n_e<6.2\times10^4\,\rm{cm^{-3}}$.
Combined with $L_{ion}$ and $\xi$,  the distance from the center black hole to this WA is further than 0.9 pc.

The large error bars of $\xi$ in the lowest ionization component make it hard to check whether $\xi$ varied.
However, based on the variation of Si L-shell lines, we conclude that the ionization state of the lowest ionization component changed over the 17 day span. 
Using the same method as above, we find that $n_e>5.5\times10^4\,\rm{cm^{-3}}$ and that the distance to this component is less than 2.5 pc.
It might be common that a lower ionization component reaches a new photoionization equilibrium easier than a higher ionization components \citep{krongold05}, as in NGC 3783, NGC 4051, NGC 5548.

%******************************************************************
\subsection{On The Origin of The Emission Lines}
The location of gas producing soft X-ray emission lines is unclear. 
These lines may originate from just the WA \citep{netzer03}, very close to or within the BLR \citep{costantini07, longinotti08, longinotti10}, within the Narrow Line Region (NLR) \citep[]{guainazzi07}, or between the BLR and the NLR \citep{nucita10}.
The soft X-ray emission lines in Mrk 290 are redshifted with respect to the systemic velocity, and the velocity did not change over 3 years, so they are probably not related to the WAs.

The FWHM of Fe k$\alpha$ is similar to the width of H$\beta$ \citep[5323.5 km s$^{-1}$;][]{vestergaard06}, indicating a possible origin in the BLR \citep{bianchi08}. 
However, forcing the lines to be narrow does not make the fits significantly worse, which suggests that the torus is still the primary origin for this fluorescent line \citep{bianchi07}.

%******************************************************************
\subsection{The Torus Origin}
With the reverberation mapping results \citep{denney10}, the black hole mass of Mrk 290 is $2.43\pm0.37\times10^7\,M_\odot$.
The time delay of H$\beta$, $8.72^{+1.21}_{-1.02}$ days, indicates that the radius of the BLR is 0.007 pc. 
The thermal bolometric luminosity calculated from the SED, $L_{bol}=2.5\times10^{44}\,\rm{erg\,s^{-1}}$, gives the radius of the torus: 1.1pc \citep[e.g.][]{krolik01}.
Thus, the radius of the torus seems similar to the distance of the WAs.

The thermal photoionization equilibrium curve \citep{krolik81} is presented in Fig.~\ref{fig:Scv} for the high-state SED.
The curve marks the points of thermal equilibria in $T$ - $\Xi$ plane, where $T$ is the electron temperature and the ``pressure" ionization parameter $\Xi=\xi/4\pi ckT$.
The three WAs best modeled in the high-state spectrum are roughly in pressure balance as shown in Fig.~\ref{fig:Scv}, indicating that all three components may occupy the same spatial region. 
Combined with the distance, we suggest that the origin of the WAs is a thermal wind from the torus, the model proposed by \citet{krolik01}.

The WAs in the RGS spectrum do not show significantly different ionization parameters and column densities compared to that found from the HETGS spectra.
This supports the model that the WAs come from the torus.
The large change of the outflow velocity, from $450\,\rm{km\,s^{-1}}$ to $1260\,\rm{km\,s^{-1}}$, gives a large acceleration $\rm{0.6\,cm\,s^{-2}}$.
Radiation pressure is an attractive means for accelerating the gas \citep{crenshaw03}, and the acceleration can be written as
$a=\frac{\kappa L}{4\pi r^{2}c}-\frac{GM}{r}$, where $\kappa$ is the absorption cross-section per unit mass. 
For Mrk 290, the acceleration by radiation pressure is $\sim\rm{0.01\,cm\,s^{-2}}$, much less than the observed one.
As a result, an entirely new WA gas may have passed through our line of sight during the \xmm observation.

A natural geometry for WA winds is bi-conical \citep{dorodnitsyn08}.
Assuming this, the mass loss rate can be estimated.
We use a formula for the mass outflow rate derived by \citet{krongold07}:
\begin{equation}
\dot{M}_{out} = 0.8 \pi m_p N_H v_r R f(\delta,\phi)
\end{equation}
where $f(\delta,\phi)$ is a factor that depends on the particular
orientation of the disk and the wind, and for all reasonable angles
($\delta> 20^o$ and $\phi>45^o$) is of order of unity.
With the observed $v_r$ (the line of sight outflow velocity), $N_H$, and the distance of torus $R=1.1\,\rm{pc}$, the mass outflow rate
contributed by the medium and lowest ionized WAs is $\sim1M_{\odot}$ per year, of the same order as the values in the simulation \citep{dorodnitsyn08}.
The outflow mass rate estimate is larger than the accretion rate, $\dot{M}_{acc}=\frac{L_{bol}}{c^2\eta}\sim0.5M_{\odot}$ per year, assuming $\eta=0.1$.

The escape velocity ($v_{esc}=\sqrt{2GM/R}$) from the location of the torus is about $440\,\rm{km\,s^{-1}}$, which is nearly identical to the outflow velocities of the WAs in Mrk 290.
The outflow velocity is only a lower limit to the actual speed, thus the warm gas would not likely fall back, and would move 5 kpc away from the central region into the host galaxy over the AGN lifetime ($10^7$ yr).
The total mass lost due to the WAs is $M_{out}\sim10^7\,M_{\odot}$, and the integrated kinetic energy (given by $M_{out}v^2/2$) is $\sim2\times10^{55}\,\rm{ergs}$ in $10^7$ yr. As suggested by \citet{krongold07}, the hot interstellar medium in the host galaxy can be heated to $10^7$ K and would evaporate, ceasing star formation.
As a consequence, this outflow mass rate may affect the evolution of the host galaxy and the surrounding intergalactic medium, over the AGN lifetime.

%******************************************************************
\subsection{The Connection between UV and X-ray Absorbers}
Mrk 290 was observed by \fuse simultaneously with \chandra.
\citet{kaiser06} detected three intrinsic absorbers by the \ovi~doublet in the UV band, blueshifted by 526, 470 and $\rm{439\,km\;s^{-1}}$ with respect to the systemic velocity.
Comparing them with the outflow velocities of the three WAs in X-ray band, $540\pm150$, $450\pm30$ and $\rm{390\pm60\,km\,s^{-1}}$, we find the UV absorption components correspond to the X-ray WAs within the uncertainties.
One geometry to interpret this connection is that high-density, low-ionization UV-absorbing clouds are embedded in a low-density, high-ionization material that dominates the X-ray warm absorber \citep{kriss04}.
\citet{kaiser06} estimated that the hydrogen column density is $9\times10^{14}\,\rm{cm^{-2}}$. 
Assuming $n_e>6\times10^4\,\rm{cm^{-3}}$, the thickness of the UV-absorbing clouds must be less than $1.5\times10^5\,\rm{km}$.
In fact, many X-ray observed absorption components appear to be accompanied by corresponding UV absorption components with the same outflow velocity \citep[e.g.][]{steenbrugge09}.

%%%%%%%%%%%%%%%%%%%%%%%%%%%%%%%%%%%%%%%%%%%%%%%%%%%%%%%%%%%% 
\section{Conclusions}
We summarize the main results of this analysis of the Mrk 290 high resolution X-ray spectra as follows:

1. The \chandra-HETGS spectra reveal complex absorption features including the Fe M-shell unresolved transition array (UTA), indicating the presence of intervening ionized absorbing gas.
Photoionization modeling shows that a combination of three ionization absorbers with similar kinematics (outflow velocities $\sim\,450\,\rm{km\,s^{-1}}$) can fit the data and are consistent with the three absorption components in the previous far UV band study. The ionized component with highest ionization is required by the presence of several absorption lines of \fexxib, which are rarely observed.

2. The ionizing continuum increased 40\% during the \chandra observation, on a timescale of 17 days. 
Using the lack of response in the opacity of the higher ionization component to this fluctuation, we constrain the distance from the ionizing source to  $>$ 0.9 pc .
Si L-shell line ratios show significant changes and are related to the UTA, so they result from same gas with the ionization level. 
If the variation derived from the Si L-shell lines of the lowest ionization component is real, we may set an upper limit of 2.5 pc for the distance to the central source. 
The three ionization components lie on the stable branch of the S-curve, showing roughly the same gas pressure, which indicates that they occupy the same spatial region.
We conclude that the warm absorbing gas originates in the thermal wind from the torus.

3. During the \xmm observation, the ionizing luminosity of Mrk 290 was only half of the high luminosity of the \chandra observation.
The RGS spectrum is well fitted by a two phase warm absorber, with several additional local absorption lines attributed to the high velocity cloud Complex C.
Neither the ionization parameter $\xi$ nor the column density $\rm{N_H}$ of the two absorbing components varied significantly, compared to the results from \chandra observations. 
The outflow velocities of both components increased by a factor of 2.5.
However, the radiation pressure in Mrk 290 is not enough to accelerate the gas to $\sim\,1260\,\rm{km\,s^{-1}}$.
We suggest that entirely new WA gas from the torus passed through our line of sight.

4. The estimated mass outflow rate of the ionized wind is $\sim1M_{\odot}$ per year, larger than the nuclear accretion rate $\sim0.5M_{\odot}$ per year, and the input kinetic energy is $\sim2\times10^{55}\,\rm{ergs}$.

5. In both high- and low-state spectra the \fek$\alpha$ emission lines are detected.
The \ovii~and \neix~forbidden lines are the most prominent soft X-ray emission lines, with a mean redshift velocity of $\sim\,700\,\mathrm{km\,s^{-1}}$ with respect to the systematic velocity. 
There seems to be no relation between emission lines and warm absorbers.

%%%%%%%%%%%%%%%%%%%%%%%%%%%%%%%%%%%%%%%%%%%%%%%%%%%%%%%%%%%% 

\section*{acknowledgements}
We thank the help from the scientists of the MIT Kavli Institute, and Tim Kallman for detailed instructions about XSTAR.
We thank the anonymous referee for providing useful comments on our work.
Support for this work was provided by the National Aeronautic Space Administration through the Smithonian Astrophysics of Observation contract SV3-73016 to MIT for support of the Chandra X-ray Center, which is operated by SAO for and on behalf of NASA under contract NAS8-03060,
by Program for New Century Excellent Talents in University (NCET), and by Chandra grant GO8-9113B.
Also our work is supported under the National Natural Science Foundation of China under grants (10878010, 10221001 and10633040) 
and the National Basic Research Program (973 programme no. 2007CB815405).

\clearpage

%%%%%%%%%%%%%%%%%%%%%%%%figures here

\begin{figure*} %  figure placement: here, top, bottom, or page
   \centering
   \includegraphics[angle=-90,width=6in]{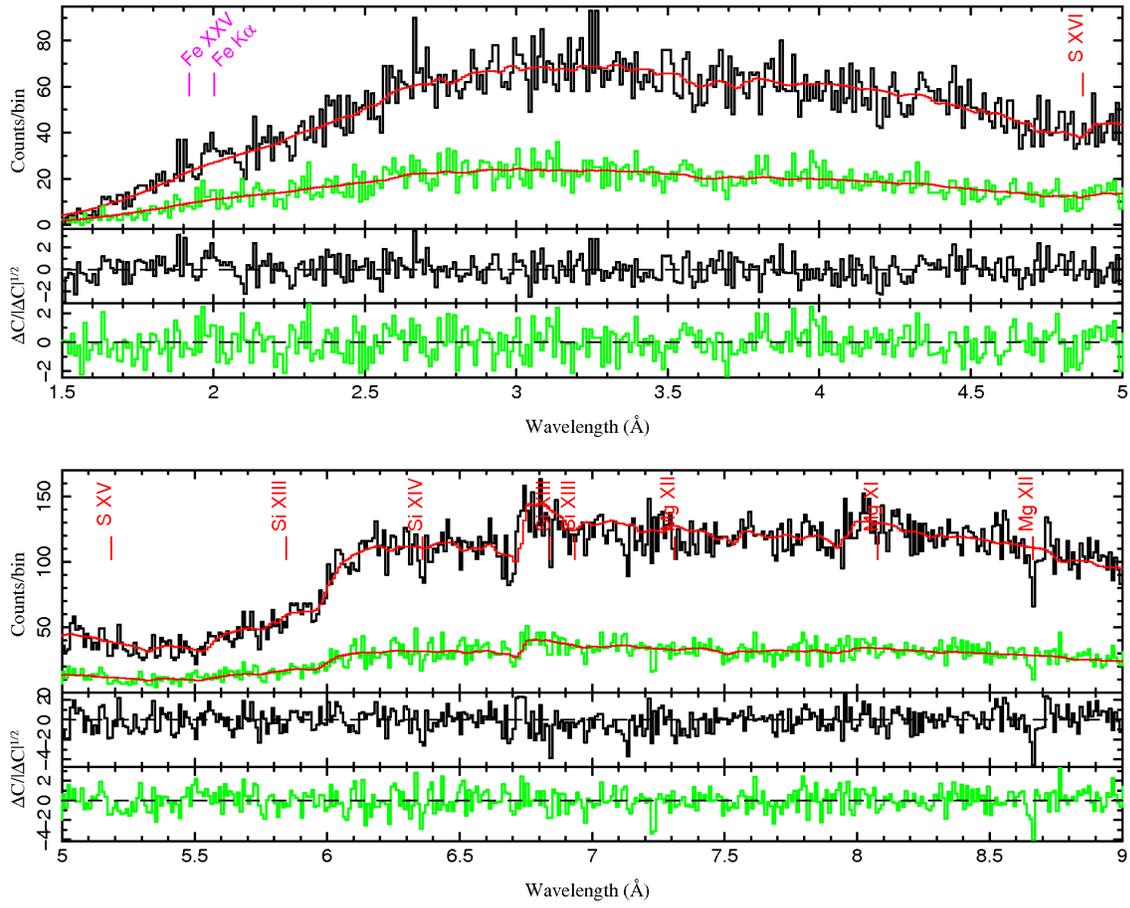} 
   \caption{Combined MEG and HEG 1st order spectra compared to the best-fitting power-law plus black body continuum model. Residuals are given in the bottom two panels for the high-state (black line) and for the mid-state (green line).
   The data were binned to 0.01 \AA. Possible spectral features are marked. In the residuals, $C$ denotes for the Cash Statistic, and $\sqrt{\Delta C}$ is a proxy for $\chi$ because $\Delta C$ is distributed as $\chi^{2}$.}
   \label{fig:none1}
\end{figure*}

\setcounter{figure}{1}

\begin{figure*} %  figure placement: here, top, bottom, or page
   \centering
   \vskip -0.6in
   \includegraphics[angle=-90,width=5.5in]{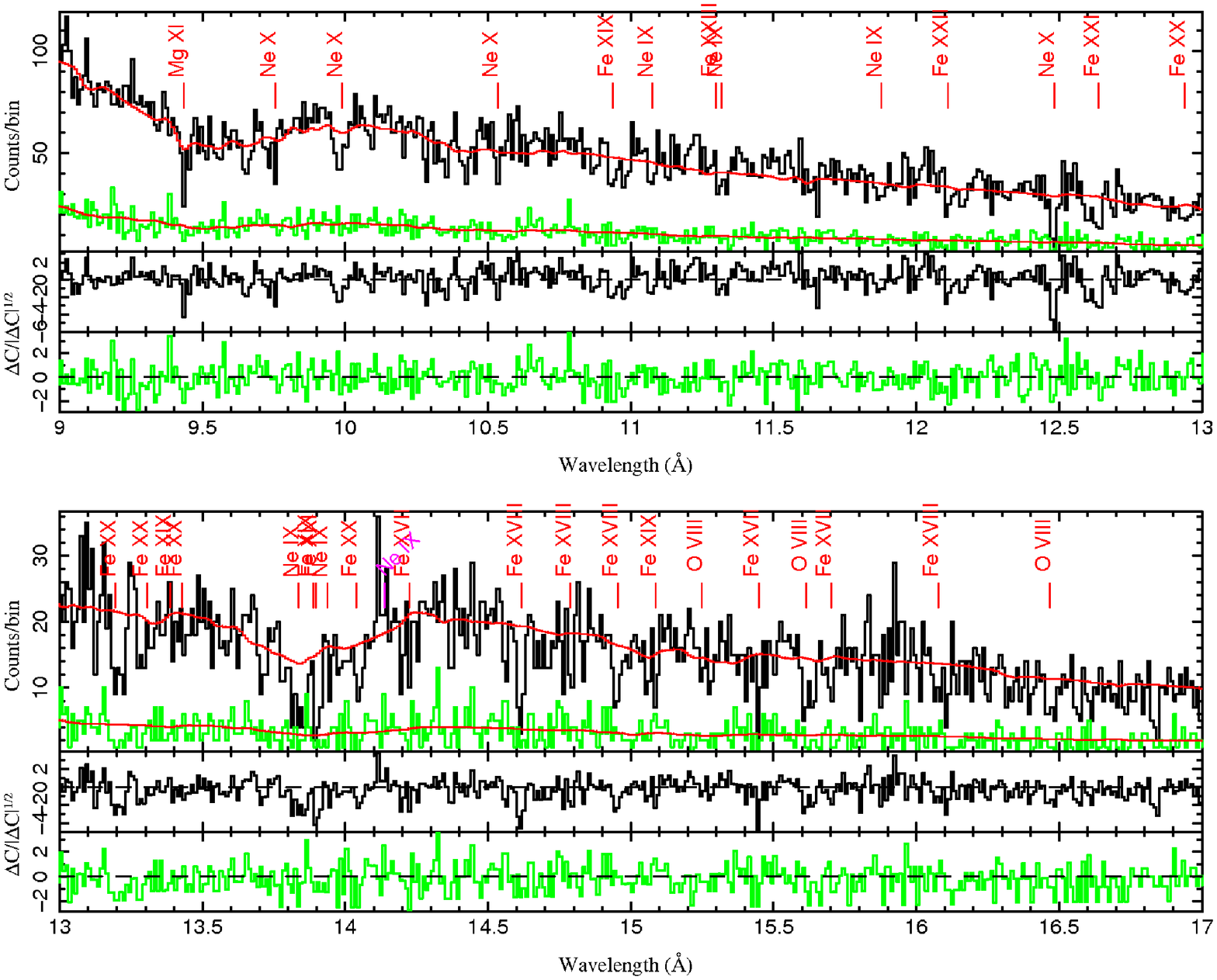} 
      \includegraphics[angle=-90,width=5.5in]{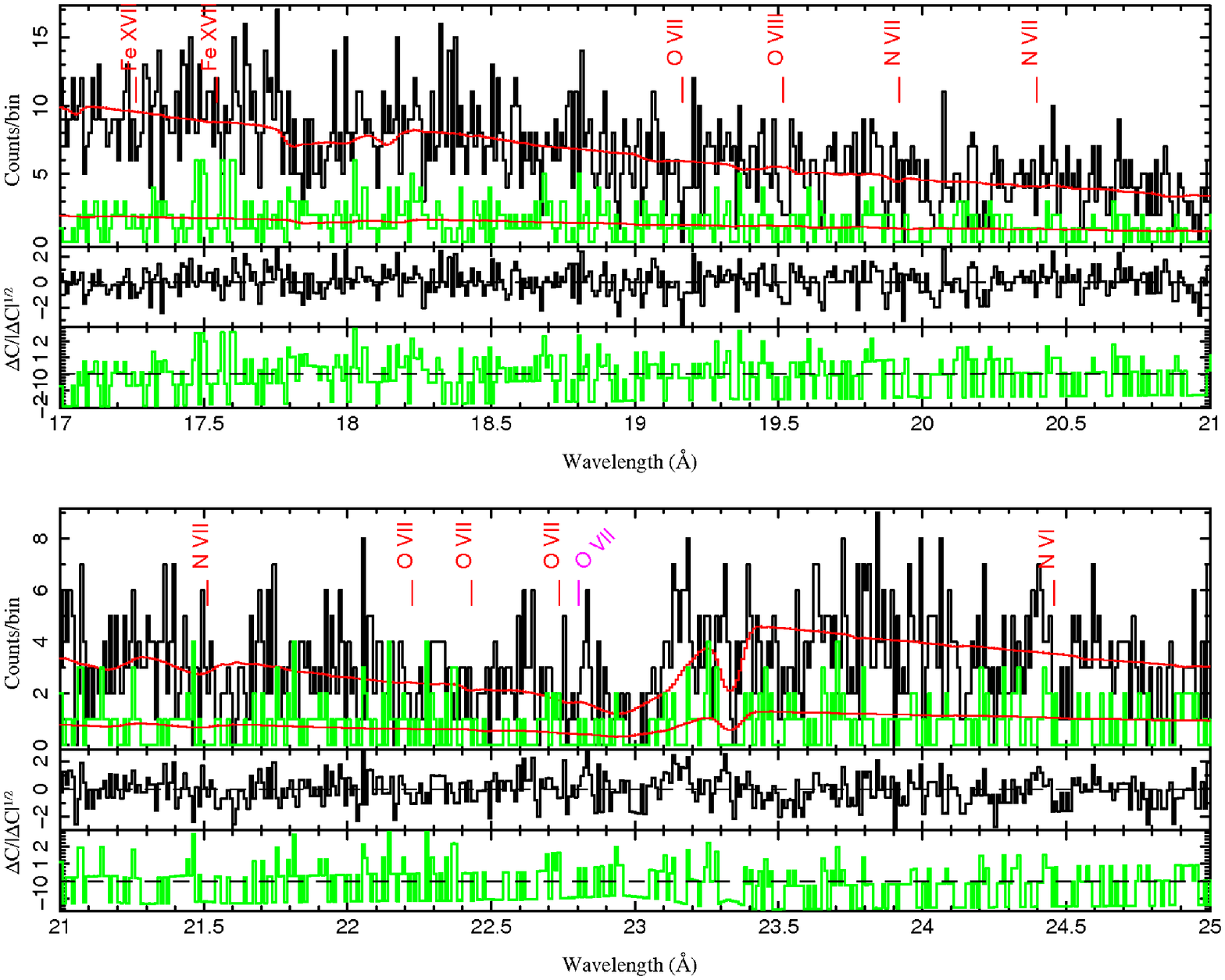} 
   \caption{{\it Continued.}}
   \label{fig:none2}
\end{figure*}

\begin{figure} %  figure placement: here, top, bottom, or page
   \centering
   \includegraphics[angle=-90,width=6in]{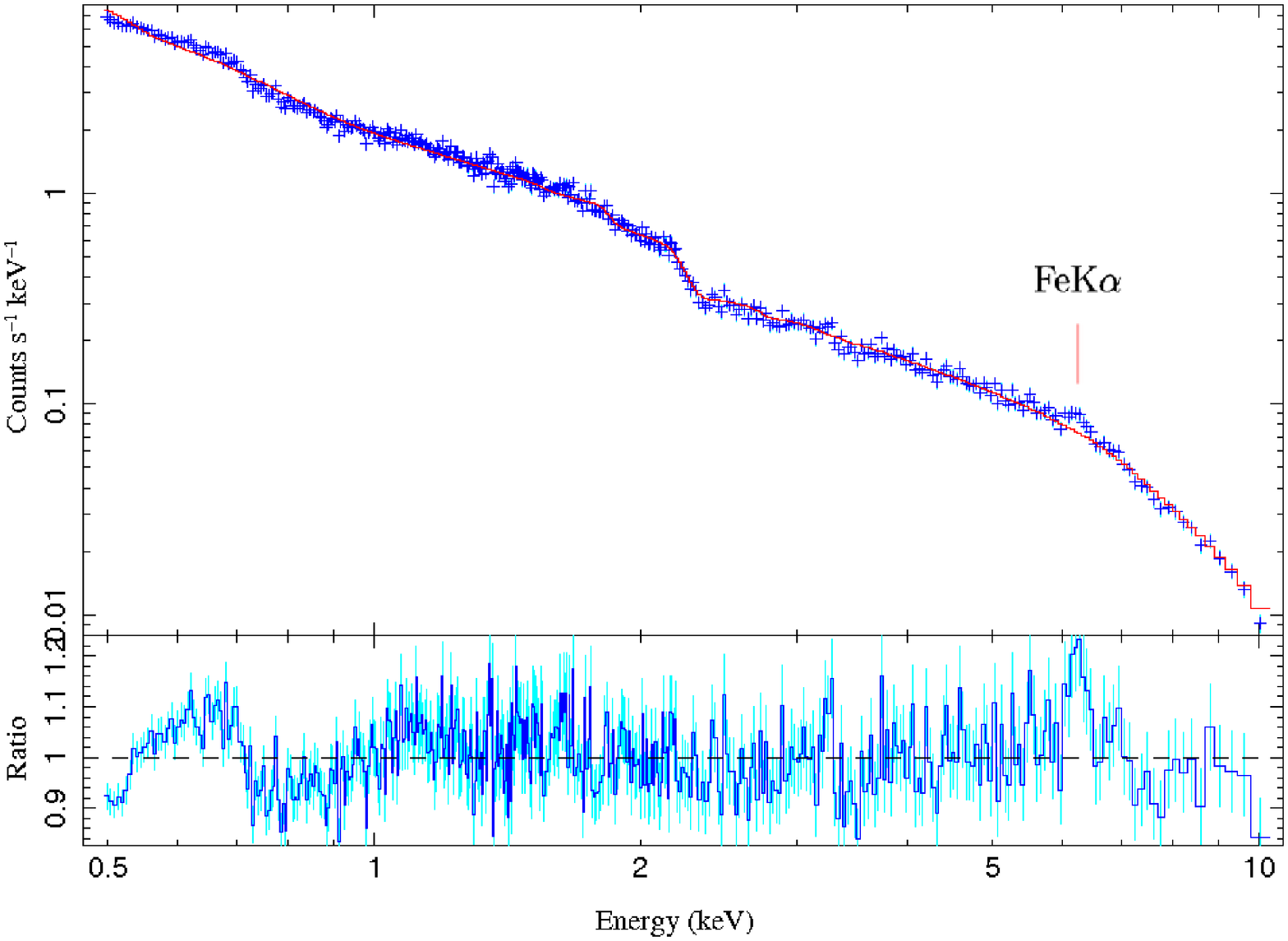} 
   \caption{Comparison of the \xmm {\it pn} spectrum to the best-fitting power-law plus black body model in the 0.5-10 keV band. Iron K$\alpha$ emission and  an absorbing feature from 0.7 to 0.9 keV can be seen clearly in the ratio of the data to the model. The absorption is due to the warm absorber.}
   \label{fig:pn}
\end{figure}

\begin{figure} %  figure placement: here, top, bottom, or page
   \centering
   \includegraphics[angle=-90,width=6in]{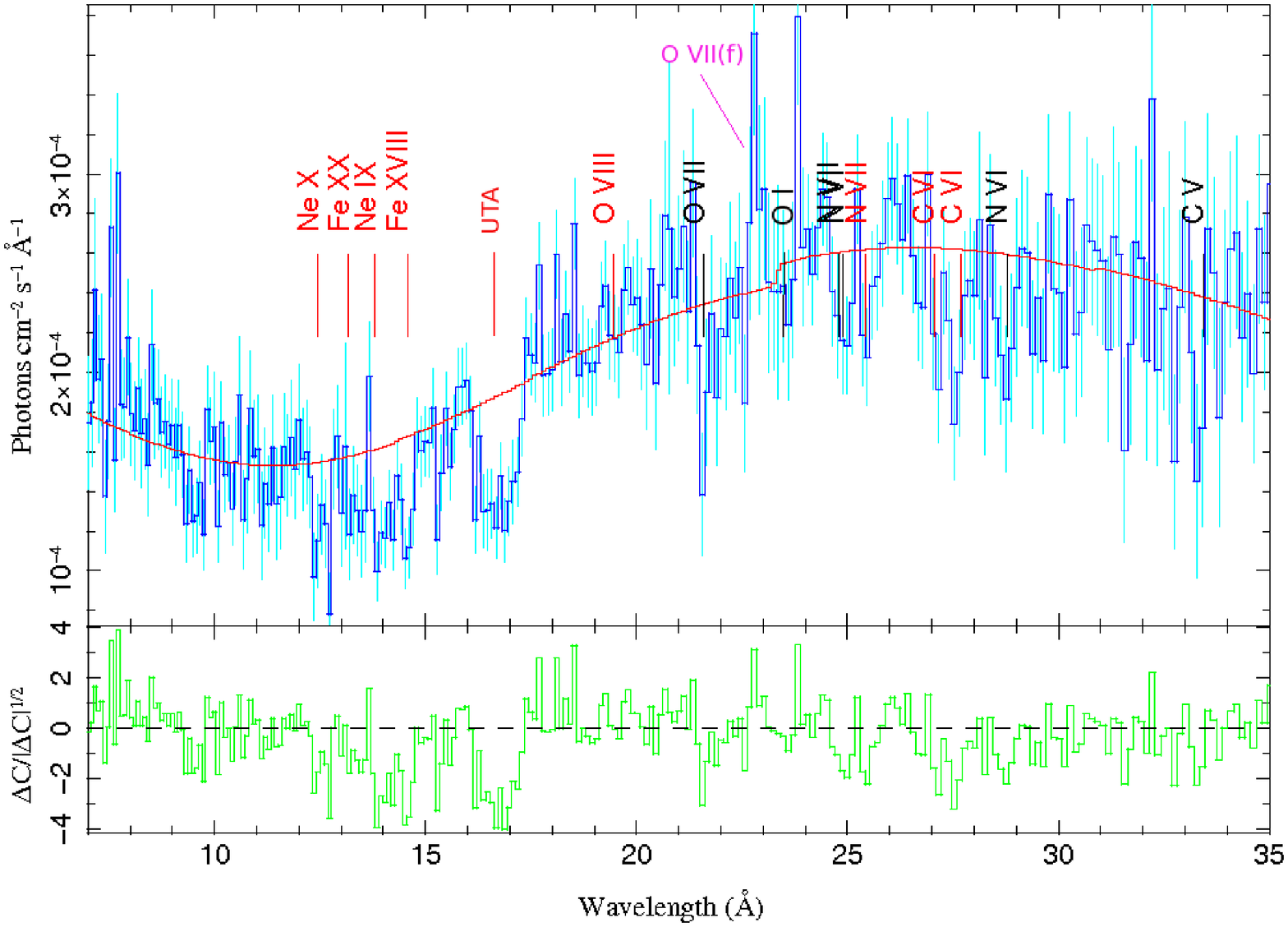} 
   \caption{Comparison of the RGS spectrum to best-fitting power-law plus black body model from {\it pn} spectral fit. The Fe UTA is clearly observed in the 16.0 - 17.5 \AA~range, along with other features that are marked.}
   \label{fig:rgsc}
\end{figure}

\begin{figure} %  figure placement: here, top, bottom, or page
   \centering 
   \includegraphics[angle=0,width=3.0in]{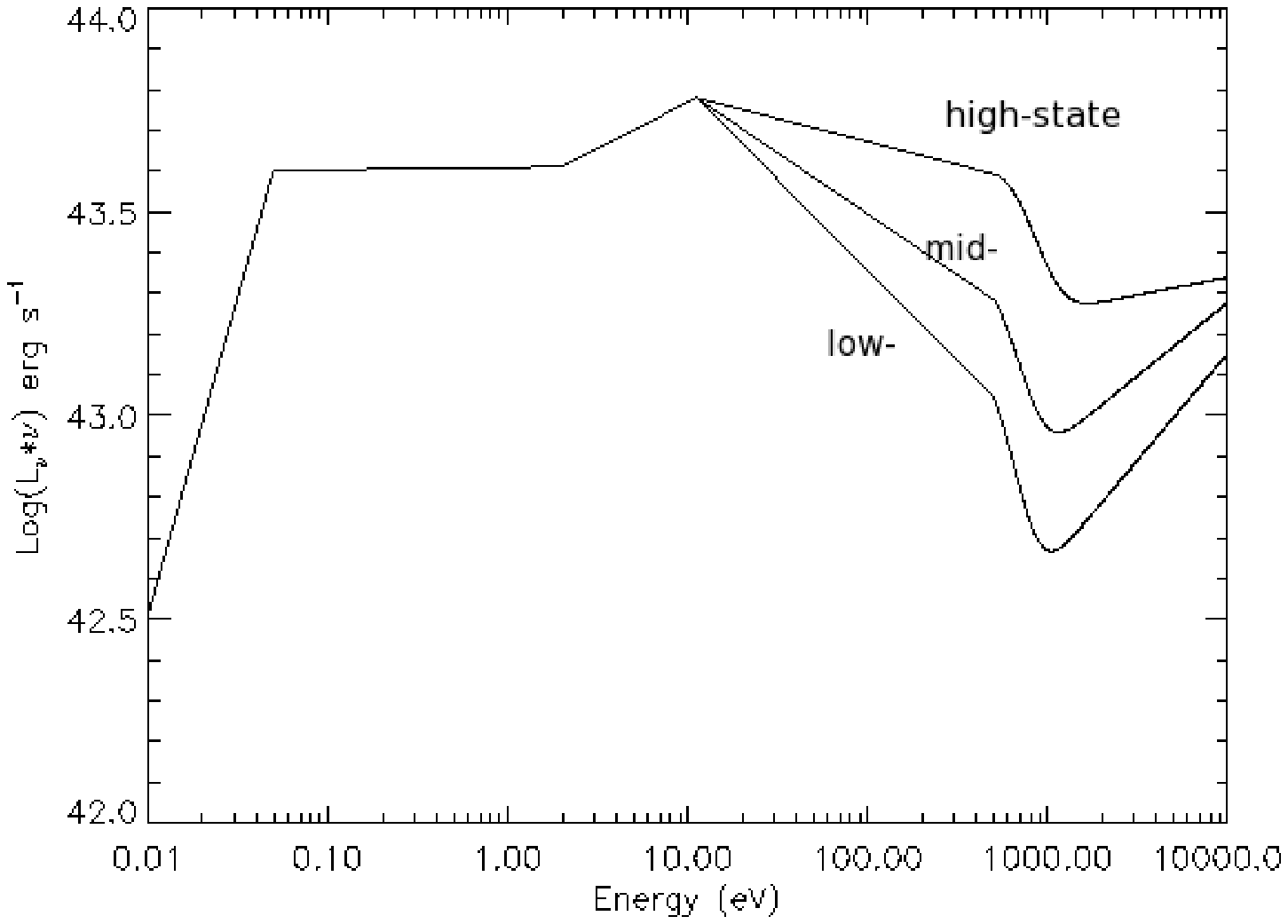} 
   \caption{The spectral energy distributions of high-, mid- and low-states used for photoionization modeling. }
   \label{fig:sed}
\end{figure}

\begin{figure*} %  figure placement: here, top, bottom, or page
   \centering
      \vskip -0.6in
   \includegraphics[angle=-90,width=6in]{fig6a.eps} 
      \includegraphics[angle=-90,width=6in]{fig6b.eps} 
   \caption{The photoionization models for coarsely binned HETGS spectra. The upper panel shows the high-state spectrum and the lower panel shows the mid-state spectrum. In the residuals, the green lines indicate the model with one WA component, while the red lines represent the model with two WA components. The difference between the two models is primarily found at the Fe UTA feature. The residuals around 23 \AA~are partly due to the \ovii(f) emission line.}
   \label{fig:twohigh}
\end{figure*}

\begin{figure*} %  figure placement: here, top, bottom, or page
   \centering
   \includegraphics[angle=-90,width=6.8in]{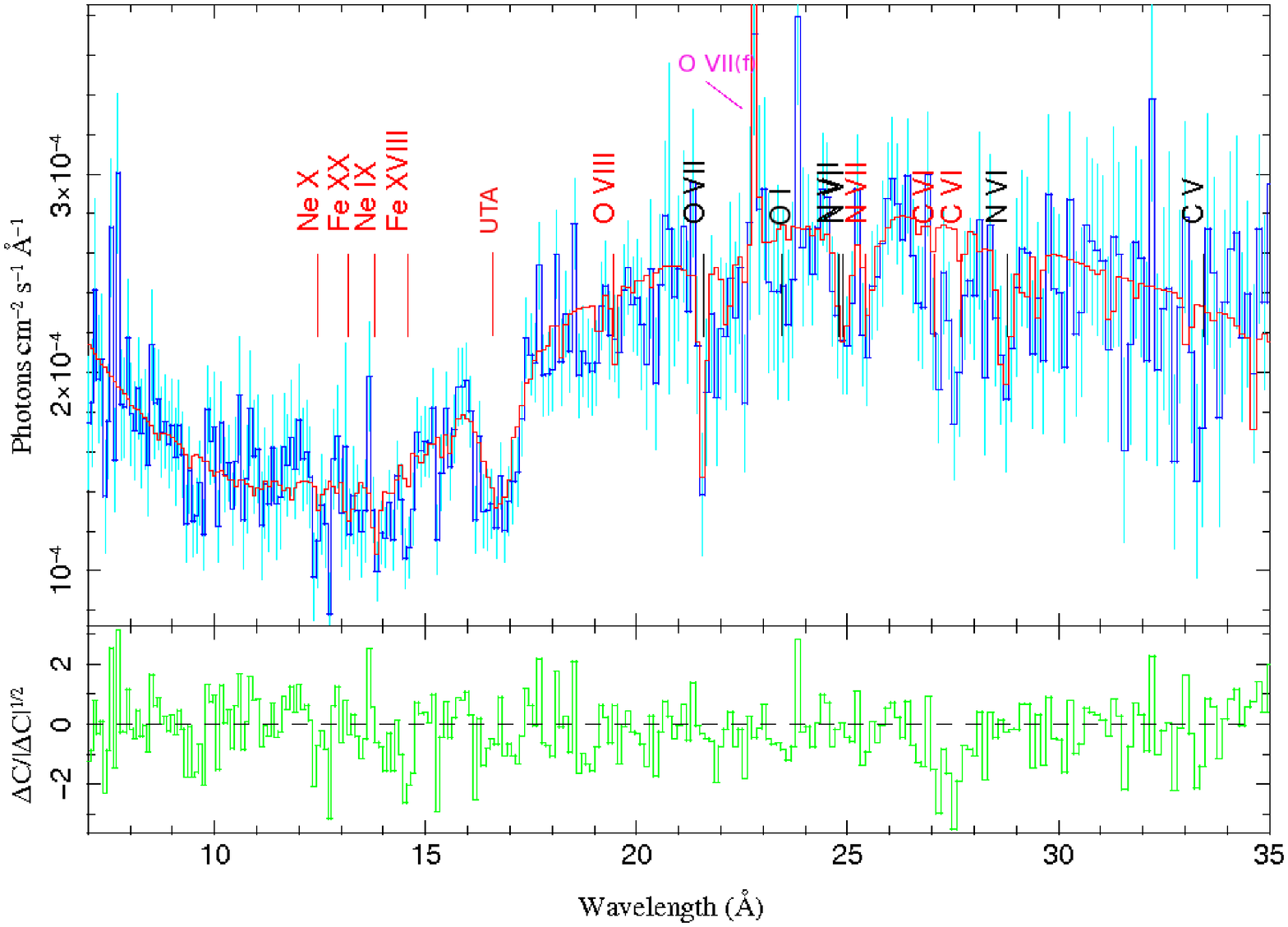} 
   \caption{The photoionization model for the low-state RGS spectrum. The lines labeled in black are at $z=0$ and are due to Galactic gas. The \ovii(f) emission line is significant.}
   \label{fig:rgsflux}
\end{figure*}

\begin{figure*} %  figure placement: here, top, bottom, or page
   \centering
   \includegraphics[angle=-90,width=6in]{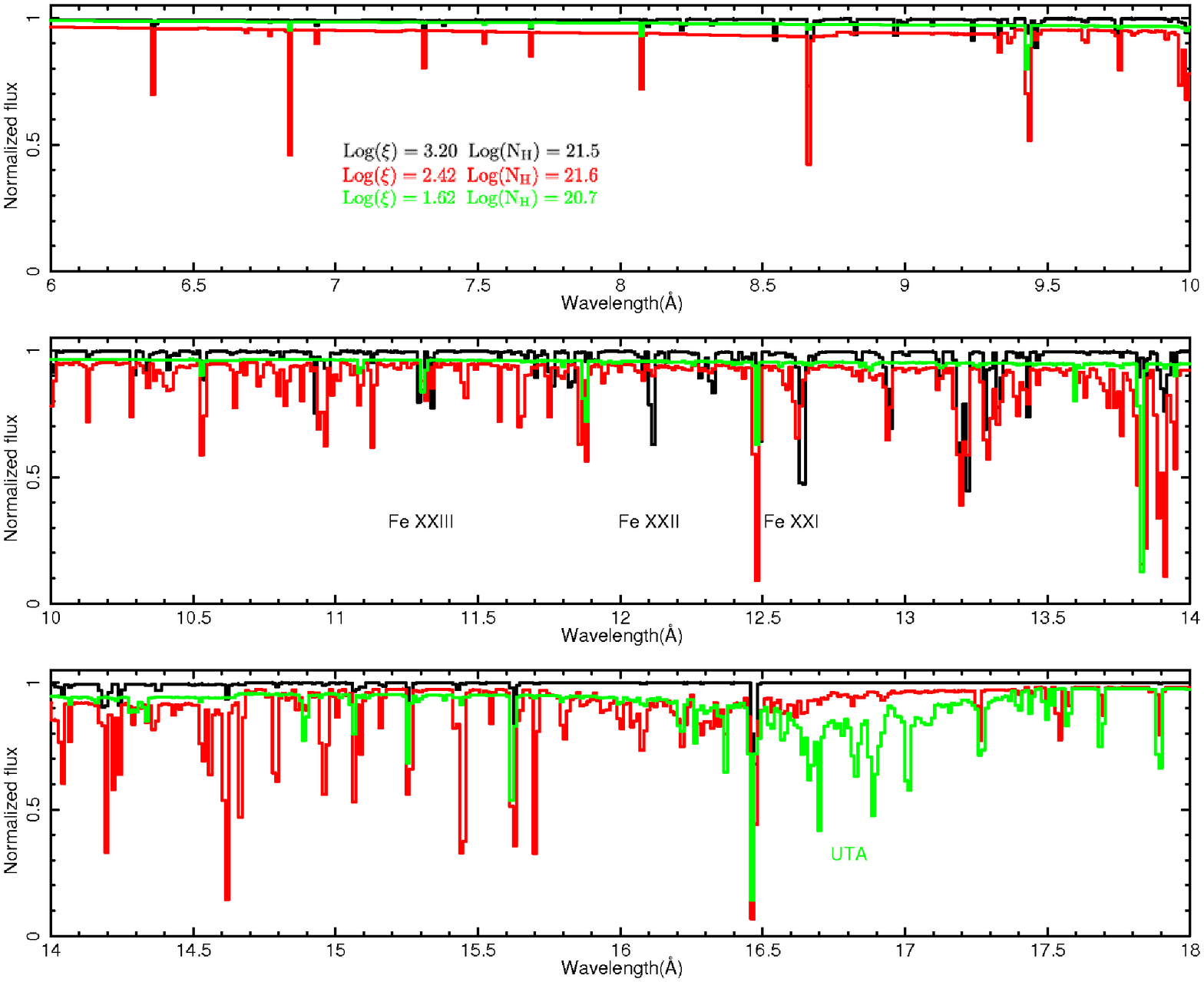} 
   \caption{Three WA components are used to model the high-state spectrum of Mrk 290, shown on a scale where 1 is the incident continuum level. 
   The three lines of \fexxib~marked in black indicate the highest ionized component, while the broad Fe UTA results from the lowest ionization component (in green).}
   \label{fig:model}
\end{figure*}

\begin{figure*} %  figure placement: here, top, bottom, or page
   \centering
   \includegraphics[angle=0,width=6in]{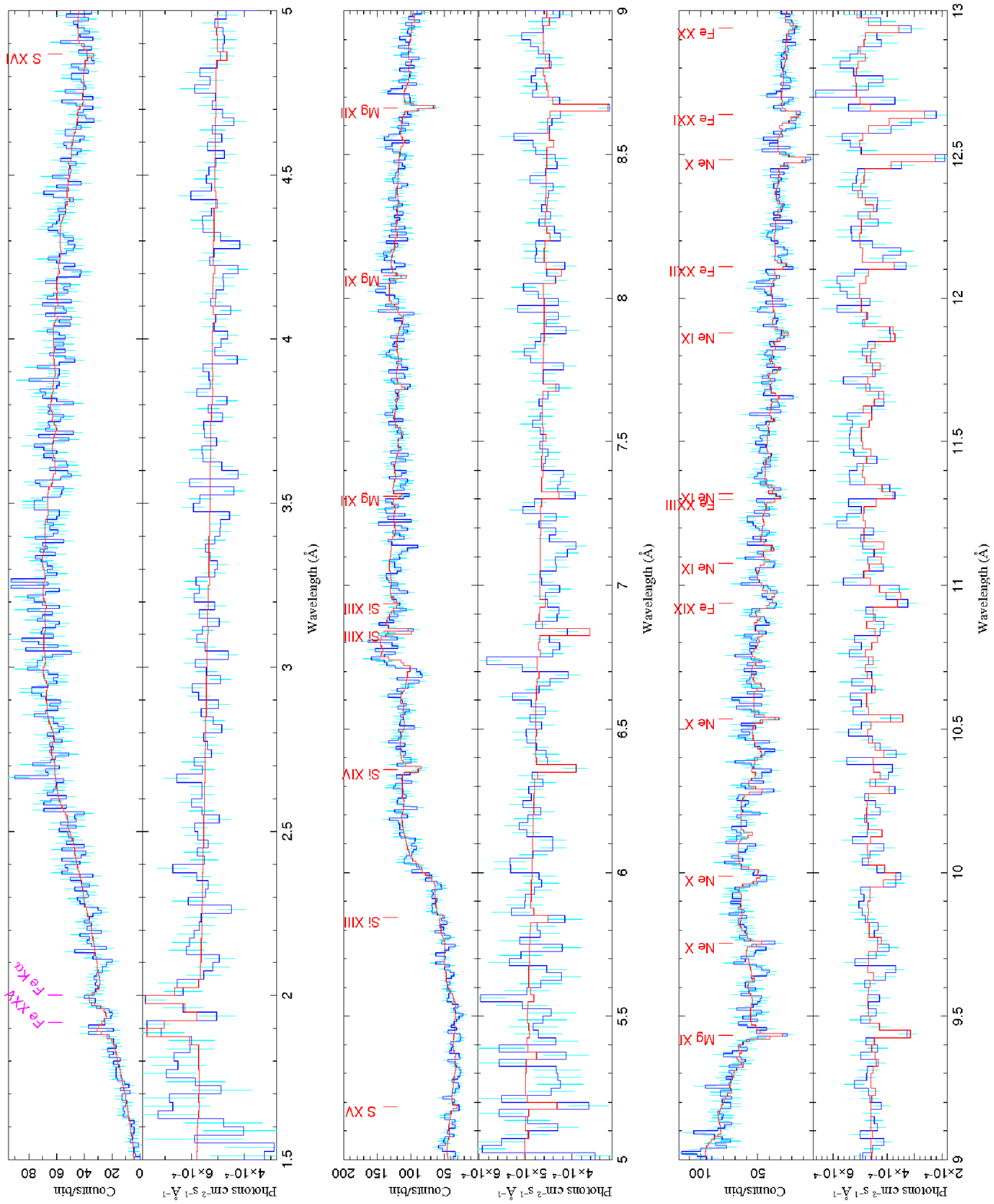} 
  \caption{The best fit model for high-state spectra in the 1.5-25 \AA~band, which consists of three WA components. Four fitted emission lines are marked in purple. The upper part and lower part of each panel show the count spectra and the flux spectra, respectively.}
   \label{fig:spec1}
\end{figure*}

\setcounter{figure}{8}

\begin{figure*} %  figure placement: here, top, bottom, or page
   \centering
   \includegraphics[angle=0,width=6in]{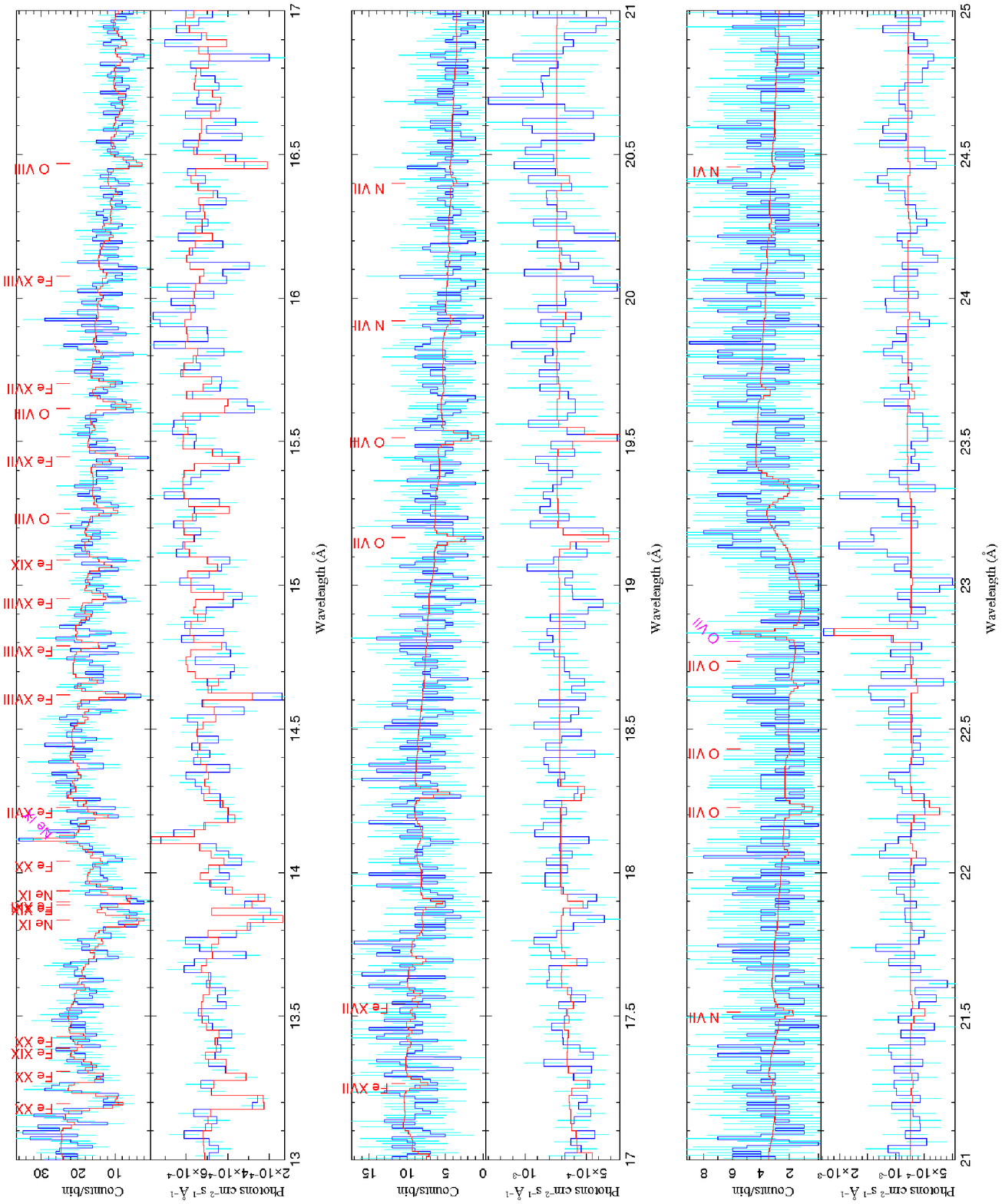} 
  \caption{{\it Continued.}}
   \label{fig:spec2}
\end{figure*}

\begin{figure*} %  figure placement: here, top, bottom, or page
   \centering
   \includegraphics[angle=0,width=6in]{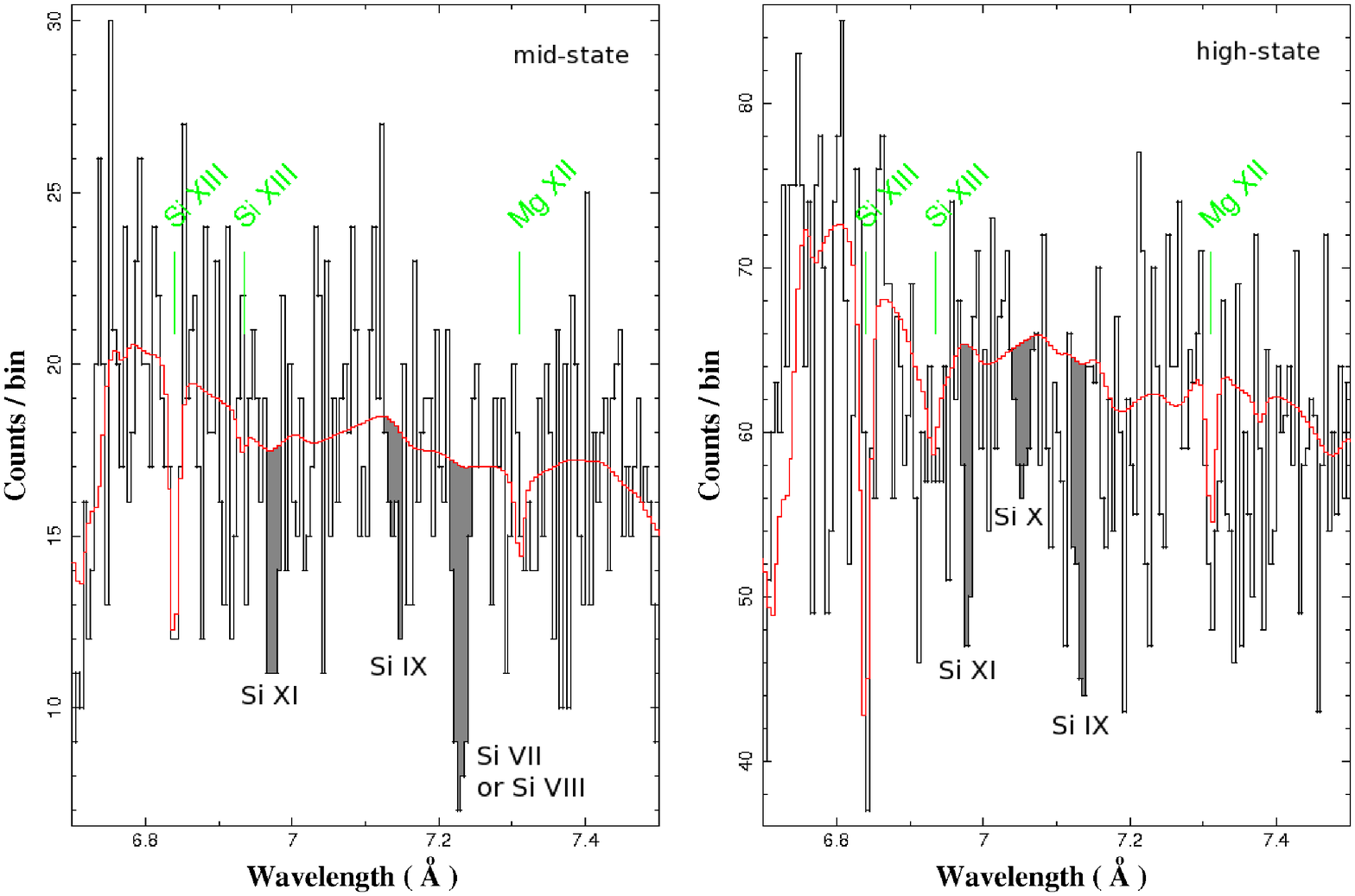} 
   \caption{The 6.7-7.5 \AA~regions of mid- and high-state spectra are shown in the left and right panel, respectively. The \sivii/\siviii~line at 7.23 \AA~is significantly stronger in the mid-state than in the high state, indicating that the ionization state of the WA responded to the continuum changes.}
   \label{fig:silicon}
\end{figure*}

\begin{figure} %  figure placement: here, top, bottom, or page
   \centering
   \includegraphics[angle=0,width=3in]{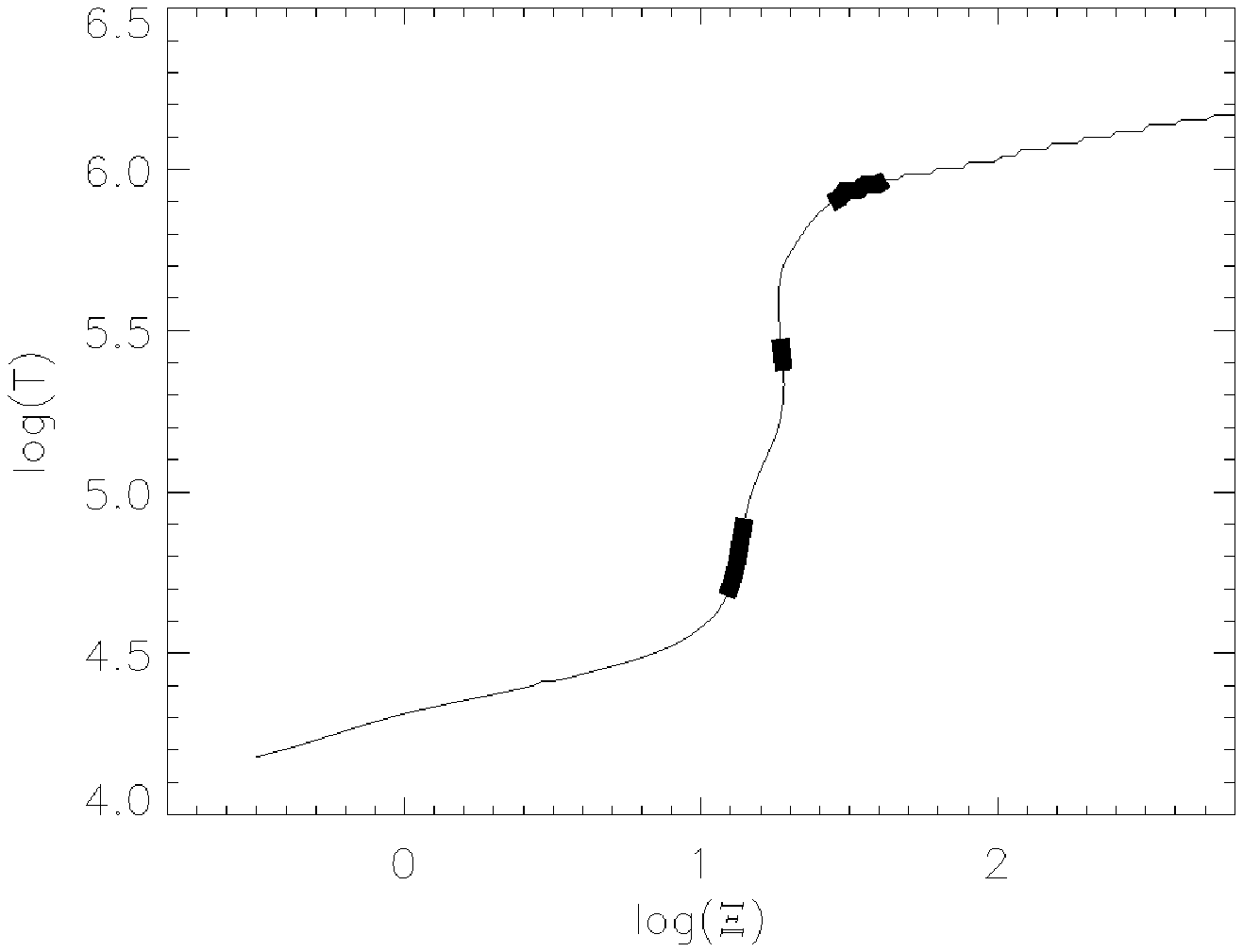} 
   \caption{Thermal stability curve for a low-density gas illuminated by the high-state continuum of Mrk 290. The quantity $\Xi=\xi/4\pi ckT$ is the pressure ionization parameter. The black sections show the locations of the three warm absorbers, with the corresponding uncertainties on the ionization parameters. The similar value of $\Xi$ indicates that the three components have roughly the same gas pressure.}
   \label{fig:Scv}
\end{figure}
 
\clearpage
%%%%%%%%%%%%%%%%%%%%%%%%tables here

\begin{table*}
\centering
  \caption[]{ \chandra HETGS observation log}
   \label{tab:id}
   \begin{tabular}{cccc}
   \hline\noalign{\smallskip}
ID &  Start Time  & Exposure time \\
   \hline\noalign{\smallskip}
4399 & 2003.06.29 01:38:04 & 85.1 ks \\
3567 & 2003.07.15 01:18:47 & 55.1 ks \\
4441 & 2003.07.15 22:54:28 & 60.9 ks \\
4442 & 2003.07.17 11:18:15 & 50.2 ks \\
   \noalign{\smallskip}\hline
   \end{tabular}
\end{table*}

\begin{table*}
  \begin{center}
  \caption[]{ \xmm observation log}
   \label{tab:idxmm}
   \begin{tabular}{cccc}
   \hline\noalign{\smallskip}  
   ID &  Start Time  & Filtered time \\  
   \hline\noalign{\smallskip}
0400360201 & 2006.04.30 16:47:58 & 20 ks  \\ 
0400360301 & 2006.05.02 15:21:00 & 7.5 ks \\   
0400360601 & 2006.05.04 14:45:53 & 15 ks  \\  
0400360801 & 2006.05.06 14:39:11 & 14 ks  \\  
   \noalign{\smallskip}\hline
   \end{tabular}
  \end{center}
\end{table*}

\begin{table*}
  \begin{center}
  \caption[]{Continuum parameters}
   \label{tab:conti}
   \begin{tabular}{ccccccl}
   \hline\noalign{\smallskip}
Date &State & $^a\Gamma$  & $^b$kT   & $\rm{Flux_{2-10keV}}$                        & $\rm{L_{2-10keV}}$             & IDs and notes \\
         &          &                         &{\tiny (eV)} & \tiny $\rm{(10^{-11}\,erg\,s^{-1}\,cm^{-2})}$  & \tiny $\rm{(10^{43}\,erg\,s^{-1})}$  &                         \\
   \hline\noalign{\smallskip}
2003.06.29 & high- & $1.86\pm0.02$ & $81\pm6$  & 1.71 &   3.13   &  3567,4441,4442 \\   
2003.07.15 & mid- &  $1.55\pm0.02$ & $69\pm9$ & 1.20 &   2.19   & 4399 \\
2006.04.30 & low- &  $1.58\pm0.01$ & $87\pm1$  & 0.88 & 1.61 & EPIC {\it pn}, RGS \\
   \noalign{\smallskip}\hline
   \end{tabular}
  \end{center}
 
$^a$The photon index of the power-law component. 
$^b$The temperature of the black-body component.

\end{table*}

\begin{table*}
  \caption[]{Absorption lines from the high resolution spectra with Poisson probability greater than 0.995} 
  \label{Tab:habs}
  \begin{tabular}{ccccll}
  \hline\noalign{\smallskip}
$\lambda_{obs}$  & Flux & Outflow Velocity & EWs & Poisson & Ion Name \& $\lambda_{rest}$      \\
 \tiny{(\AA)} & \tiny ($10^{-6}\,\mathrm{ph\,cm^{-2}\,s^{-1}})$ & \tiny ($\rm{km\,s^{-1}}$)& \tiny (m\AA) & Probability &  \tiny $\;\;\;\;\;\;\;\;$(\AA)\\
  \hline\noalign{\smallskip}
  \multicolumn{6}{c}{High-state spectrum} \\
    \noalign{\smallskip}\hline
$  4.202\pm0.006$  & $ 3.7\pm1.8$ & $ 790\pm420$  &$7\pm4$ & 0.9990 &   \sxv~(4.088) \\ 
$  6.360\pm0.002$ & $ 2.6\pm0.5$ & $ 530\pm90$  &$5\pm1$ & 0.9995 &\sixiv~Ly$\alpha$ (6.183)  \\   
$  6.844\pm0.003$ & $ 2.5\pm0.5$ & $  270\pm110$  & $  5\pm1$ & 0.9993 & \sixiii(r) (6.648)  \\
$  7.133\pm0.001$ & $ 2.3\pm0.5$ & $380\pm60$ & $5\pm1$ & 1.0000 & \siix~(6.923,6.939) \\
$  8.664\pm0.003$ & $ 3.8\pm1.4$ & $ 440\pm120$ & $8\pm3$ &1.0000 & \mgxii~Ly$\alpha$ (8.421) \\
$  9.433\pm0.002$ & $ 3.4\pm1.0$ & $ 480\pm80$  & $  8\pm2$ & 0.9975 & \mgxi~(9.169)  \\
$  9.979\pm0.005$ & $ 3.9\pm1.4$ & $ 520\pm170$&  $ 10\pm3$ &0.9996 & \nex~Ly$\gamma$ (9.708)  \\
$10.931\pm0.008$ & $3.8\pm1.8$ & $630\pm230$ & $8\pm4$ & 0.9969 & \fexix~(10.630) \\
$12.109\pm0.005$ & $4.2\pm2.0$ & $470\pm110$  & $9\pm4$ & 0.9988 & \fexxii~(11.770) \\
$ 12.482\pm0.003$ & $ 10.0\pm2.4$ & $520\pm70$& $ 21\pm7$ &1.0000 & \nex~Ly$\alpha$ (12.134)  \\
$ 12.632\pm0.007$ & $ 6.8\pm2.8$ & $620\pm170$  &$ 17\pm7$ &1.0000 & \fexxi~(12.284)  \\ 
$ 13.205\pm0.010$ & $ 9.5\pm4.6$ & $450\pm240$  &$ 23\pm11$ &0.9999 &  \fexx~(12.824,12.846,12.864) \\
$ 13.841\pm0.010$ & $21.0\pm7.1$ & $330\pm220$ &$ 45\pm15$ &1.0000 & \neix~(13.447), \fexix~(13.423,13.375) \\
$ 13.900\pm0.001$  & $8.8\pm2.0$ &  $400\pm30$ &  $19\pm4$ & 1.0000   &  \fexix~(13.497,13.518)  \\
$ 13.902\pm0.008$ & $ 8.9\pm5.5$ & $350\pm180$ &$ 21\pm13$ &1.0000 & \fexxi~(13.507)  \\
$ 14.613\pm0.006$ & $10.1\pm4.6$ & $ 570\pm130$  &$ 23\pm11$ &1.0000 & \fexviii~(14.208)  \\
$ 14.950\pm0.015$ & $ 5.1\pm4.2$ & $ 530\pm300$  &$ 12\pm10$ & 0.9998 & \fexviii~(14.534) \\
$ 15.446\pm0.006$ & $ 7.4\pm4.7$ & $ 490\pm120$   &$ 16\pm10$ & 1.0000 &\fexvii~(15.014)  \\
$15.621\pm0.015$   &  $12.1\pm6.3$ &  $330\pm300$  &$ 24\pm12$ & 0.9999 & \oviii~Ly$\gamma$ (15.176) \\
$16.472\pm0.017$   &  $8.5\pm8.2$ & $380\pm320$  & $17\pm16$ & 0.9950 & \oviii~Ly$\beta$ (16.006) \\
$ 16.841\pm0.006$ & $ 9.2\pm3.6$ & $310\pm110$ &$ 18\pm8$ &1.0000 & \fex~(16.360)  \\
$ 18.942\pm0.006$  & $14.5\pm4.9$ & $-400\pm90$  & $24\pm8$ &0.9990 & *\oviii~Ly$\alpha$ (18.967) \\
$19.164\pm0.002$  & $23.4\pm6.2$ & $460\pm40$ &$35\pm9$ &0.9994 & \ovii~(18.627) \\
$19.519\pm0.010$  & $18.4\pm9.3$ & $390\pm170$  & $27\pm14$ &0.9950 & \oviii~Ly$\alpha$ (18.967) \\
  \noalign{\smallskip}\hline
\multicolumn{6}{c}{Mid-state spectrum} \\
    \noalign{\smallskip}\hline
$6.357\pm0.005$   & $2.1\pm1.0$  & $660\pm220$  & $8\pm4$    & 0.9983 & \sixiv~Ly$\alpha$ (6.183)  \\ 
$7.229\pm0.002$  & $2.6\pm0.7$  & - & $12\pm3$   & 1.0000 &  \siviii~(6.9990) or \sivii~(7.063) \\ 
$8.665\pm0.005$  & $2.3\pm0.7$  & $430\pm170$   & $11\pm3$   & 1.0000 & \mgxii~Ly$\alpha$ (8.421) \\
$10.940\pm0.005$ & $2.6\pm1.3$  & $370\pm140$  & $13\pm6$   & 0.9987 & \fexix~(10.630) \\
$11.348\pm0.006$ & $2.7\pm1.2$  & $440\pm170$  & $14\pm6$   & 0.9978 & \fexix~(11.029)  \\
$12.481\pm0.003$ & $8.6\pm2.5$  & $540\pm80$    & $21\pm6$   & 0.9958 & \nex~Ly$\alpha$ (12.134)  \\
$13.205\pm0.010$ & $9.5\pm4.5$  & $450\pm250$  & $23\pm11$ & 0.9986 &\fexx~(12.824,12.846,12.864)  \\
$15.624\pm0.003$ & $8.7\pm3.1$  & $260\pm50$    & $42\pm15$ & 0.9959 & \oviii~Ly$\gamma$ (15.176) \\
   \noalign{\smallskip}\hline
   \multicolumn{6}{c}{Low-state spectrum} \\
    \noalign{\smallskip}\hline
$12.396\pm0.081$  & $  8.7\pm8.0$ & $2640\pm2000$ & $62\pm57$ & 0.9984 & \nex~Ly$\alpha$ (12.134)   \\
$13.823\pm0.023$  & $11.8\pm3.3$ & $1400\pm520$   & $73\pm20$ & 1.0000 & \neix~(13.447), \fexix~(13.497,13.518)  \\
$14.550\pm0.018$  & $14.0\pm3.1$ & $1890\pm380$  & $83\pm18$ & 1.0000 & \fexviii~(14.208)  \\
$21.568\pm0.056$ & $22.1\pm11.8$ & $-490\pm780$ & $100\pm41$ & 0.9988 & *\ovii(r) (21.602) \\
$24.833\pm0.023$ & $13.8\pm6.6$ & $-100\pm280$ &$52\pm25$ &0.9981 & *\nvi~(24.898), *\nvii~Ly$\alpha$ (24.785) \\
$25.492\pm0.029$ & $11.6\pm6.6$ & $520\pm350$ & $44\pm25$ & 0.9977 & \nvii~(24.779,24.785)  \\
%$28.778\pm0.026$ & $21.4\pm9.5$ & $-100\pm270$ & $85\pm37$ & 0.9927 & *\nvi(r) (28.787) \\
   \noalign{\smallskip}\hline   
  \end{tabular}
  
  * Galactic absorption lines, the velocities of which are relative to the heliocentric system.
  
\end{table*}

\begin{table*}
  \caption[]{Emission lines in the high- and low-state spectra}
  \label{Tab:emit}
  \begin{tabular}{lcccc}
      \noalign{\smallskip}\hline
Ion & $\lambda_{rest}$ &$\lambda_{obs}$ & flux  & Outflow Velocity\\
 & \tiny{(\AA)} & \tiny{(\AA)} & \tiny{($\mathrm{10^{-6}\,ph\,cm^{-2}\,s^{-1}}$)} & \tiny{$\;\;(\rm{km\,s^{-1}}$)}\\ 
     \noalign{\smallskip}\hline
\multicolumn{5}{c}{High-state spectrum} \\
  \hline\noalign{\smallskip}
\fek$\alpha$ & 1.938  &  $1.995\pm0.008$ & $4.3\pm3.3$ & - \\ 
\fexxv &1.859  & $  1.897^{+ 0.012}_{-0.028}$ & $10.9\pm7.9$ & - \\ 
\neix(f) &13.669 & $14.115\pm0.007$ & $8.8\pm4.6$ & $-670\pm150$\\ 
\ovii(f) & 22.098 & $22.831\pm0.008$ & $54.7\pm37.3$& $-830\pm110$ \\ 
  \noalign{\smallskip}\hline
\multicolumn{5}{c}{Low-state spectrum} \\
    \noalign{\smallskip}\hline
\fek$\alpha$ &1.938   & $2.001\pm0.021$ & $4.8\pm1.7$ & - \\ 
\ovii(f) & 22.098 & $22.818\pm0.026$ & $29.0\pm16.5$ & $-660\pm350$ \\ 
    \noalign{\smallskip}\hline
  \end{tabular}
  \end{table*}

\begin{table*}
 \centering
  \caption[]{Best fit parameters for one WA component}
  \label{Tab:onewa}
  \begin{tabular}{c|ccc|cc}
  \hline\noalign{\smallskip}
State & Log$\xi$ & $N_H$ & $^a$Velocity & $C$stat/d.o.f. & $^b\Delta C$   \\
 & \tiny  (erg s $\rm{cm^{-1}}$) &\tiny  $(\rm{10^{21}\,cm^{-2}})$ & \tiny $\rm{(km\,s^{-1})}$   &  & \\
  \hline\noalign{\smallskip}
high- & $2.43\pm0.02$ & $4.9\pm0.5$ & $450\pm30$ & 2635/2344 &  431  \\
mid-  & $2.41\pm0.07$ & $4.8\pm1.3$ & $450\pm60$ &  2071/1843 &  92  \\
 \noalign{\smallskip}\hline
  \end{tabular}
  
  $^a$ The outflow velocity of WA is relative to the systematic velocity of Mrk 290.
  
  $^b$ Change in the $C$ statistics between models with and without a warm absorber.
  
\end{table*}

\begin{table*}
  \caption[]{The two phase WA parameters for three states.}
  \label{Tab:warmabs}
  \begin{tabular}{c|ccc|cccc|c}
  \hline\noalign{\smallskip}  
 State & $^a\Gamma$ & $^b$kT & L{\tiny 1-1000 Ryd} & & Log$\xi$  & $N_H$ & $^c$Velocity  & $C$stat/d.o.f. \\
 & &\tiny  (eV) &\tiny  $ \rm{(10^{43}\,erg\,s^{-1})}$  & & \tiny  (erg s $\rm{cm^{-1}}$) &\tiny  $(\rm{10^{21}\,cm^{-2}})$ &\tiny  $\rm{(km\,s^{-1})}$ & \\
  \hline\noalign{\smallskip}
high- & $1.91\pm0.01$ & $119\pm3$ & 13.7 & WA1 & $1.62\pm0.15$ & $0.54\pm0.18$ & $540\pm150$  & 414/306\\
     &   &   &  & WA2 & $2.45\pm0.04$ & $5.33\pm1.17$ & $^d450$ &  \\
  \hline\noalign{\smallskip}
$^e$mid- & $1.64\pm0.02$ & $95^{+6}_{-1}$ & 10.0 & WA1 & $1.52\pm0.55$ & 0.54 & 540  & 172/152\\
  &   &   &  & WA2 & $2.37\pm0.08$ & 5.33 & 450 &  \\
  \hline\noalign{\smallskip}
low- & $1.69\pm0.01$ & $105\pm2$ & 7.2 & WA1 & $1.57\pm0.15$ & $1.12\pm0.18$ & $1260\pm160$  & 225/183 \\
&    &    &  & WA2 & $2.50^{+0.31}_{-0.08}$ & $3.16^{+0.35}_{-1.48}$ &$1260\pm250$  &   \\
 \noalign{\smallskip}\hline
  \end{tabular}
  
  $^a$The photon index of the power-law component in the intrinsic spectra. 
  
  $^b$The temperature of the black-body component in the intrinsic spectra.
 
  $^c$ The outflow velocity of WA is relative to the systematic velocity of Mrk 290.

  $^d$ The outflow velocity is fixed to the value in Table~\ref{Tab:onewa}.
  
  $^e$ The column density and the outflow velocity of WAs are fixed to the values obtained from the high-state spectral fits.
  
\end{table*}

\begin{table*}
 \caption[]{Best fit parameters for the three WAs in the high-state spectrum}
  \label{Tab:triwa}
  \begin{tabular}{r|ccc}
  \hline\noalign{\smallskip}
 & Log$\xi$ & $N_H$ & $^a$Velocity\\
 & \tiny  (erg s $\rm{cm^{-1}}$) &\tiny  $(\rm{10^{21}\,cm^{-2}})$ & \tiny $\rm{(km\,s^{-1})}$   \\
  \hline\noalign{\smallskip}
$^b$WA1 & 1.62 & 0.54 & 540  \\
WA2 & $2.42\pm0.04$ & $4.06\pm0.62$ & $450\pm30$  \\
WA3 & $3.20\pm0.14$ & $3.49\pm1.47$ & $390\pm60$  \\
 \noalign{\smallskip}\hline
  \end{tabular}
  
  $^a$ The outflow velocity of WA is relative to the systematic velocity of Mrk 290.

  $^b$ The parameters of this lowest ionization component is fixed to the values in Table~\ref{Tab:warmabs}.
  
\end{table*}

\label{lastpage}

\end{document}